\documentclass[sigconf,nonacm]{acmart}
\usepackage[utf8]{inputenc}
\usepackage{textgreek}
\usepackage{xcolor}
\usepackage{amsmath}
\usepackage[ruled, vlined, linesnumbered]{algorithm2e}
\usepackage{bm}
\usepackage{enumitem}
\usepackage{lipsum} 
\usepackage[most]{tcolorbox}
\usepackage{tikz}
\usepackage{pgfplots}
\pgfplotsset{compat=1.18}
\usetikzlibrary{arrows.meta, positioning, backgrounds, fit}
\usepackage[aboveskip=-4pt]{subcaption}
\usepgfplotslibrary{fillbetween}
\usepackage{amsthm}
\usepackage{mdframed}
\usepackage{seqsplit}
\usepackage{placeins}
\usepackage{needspace}

\usepackage{tabularx}
\usepackage{array}
\newcolumntype{R}{>{\raggedleft\arraybackslash}X} 

\theoremstyle{definition}

\newcommand{\eat}[1]{} 

\newcommand{\ie}{{i.e.},\xspace}

\newcommand{\stitle}[1]{%
  \par
  \vspace{0.2em}%
  \noindent
  \textbf{#1.}\enspace
  \ignorespaces
}

\definecolor{hlbg}{RGB}{200,240,240}

\newcommand{\webtwo}{\texttt{\seqsplit{chatgpt-5.2-instant}}\xspace}
\newcommand{\webthree}{\texttt{\seqsplit{chatgpt-5.3-instant}}\xspace}
\newcommand{\webfour}{\texttt{\seqsplit{chatgpt-5.4-thinking}}\xspace}
\newcommand{\webfive}{\texttt{\seqsplit{chatgpt-5.5-instant}}\xspace}

\def\addlegendimage{\csname pgfplots@addlegendimage\endcsname}

\makeatletter
\newcommand\footnoteref[1]{\protected@xdef\@thefnmark{\ref{#1}}\@footnotemark}
\makeatother

\definecolor{RedPPAlt}{HTML}{DC143C}
\definecolor{Red}{HTML}{E81123}
\definecolor{Orange}{HTML}{FFB000}
\definecolor{Green}{HTML}{009E49}
\definecolor{LightBlue}{HTML}{00BCF2}
\definecolor{DeepBlue}{HTML}{001BA3}
\definecolor{Pink}{HTML}{7C3AED}
\definecolor{LightGreen}{HTML}{8BC63E}
\definecolor{Gray}{HTML}{6B7280}

\newcounter{myremark}
\renewcommand{\themyremark}{\arabic{myremark}}

\newtcolorbox{remark}[1][]{
    enhanced,
    breakable,
    before skip=10pt,
    after skip=10pt,
    colback=gray!5,
    colframe=navyblue!70!black,
    boxrule=0.3pt,
    leftrule=3pt,
    left=8pt,
    right=5pt,
    top=5pt,
    bottom=5pt,
    sharp corners,
    fonttitle=\bfseries,
    coltitle=black,
    title={Remark \refstepcounter{myremark}\themyremark #1},
    attach title to upper,
    after title={.\hspace{0.5em}},
}
\definecolor{navyblue}{RGB}{0, 0, 128}

\AtBeginDocument{%
  }

\begin{document}

\title{Understanding ChatGPT Citations: Evidence from Production Web Search Across Model Generations}

\author{Yunxuan Jiang}
\affiliation{%
  \institution{Xi'an Jiaotong University}
  \country{China}
}
\affiliation{%
  \institution{SGIT AI Lab, State Grid Corporation of China}
  \country{China}
}
\email{fengjianliu@stu.xjtu.edu.cn}

\author{Shiqi Zhang}
\affiliation{%
  \institution{National University of Singapore}
  \country{Singapore}
}
\email{sqzhang@nus.edu.sg}

\author{Anqi Xing}
\affiliation{%
  \institution{National University of Singapore}
  \country{Singapore}
}
\email{e1546677@u.nus.edu}

\author{Xiangyu Chang}
\affiliation{%
  \institution{Xi'an Jiaotong University}
  \country{China}
}
\email{xiangyuchang@xjtu.edu.cn}

\author{Xiaokui Xiao}
\affiliation{%
  \institution{National University of Singapore}
  \country{Singapore}
}
\email{xkxiao@nus.edu.sg}
\renewcommand{\shortauthors}{Jiang et al.}

\begin{abstract}
Generative engines such as ChatGPT Web retrieve webpages to answer user queries, but only a subset are shown as citations. Understanding this retrieval-to-citation process is important for generative engine optimization (GEO), which aims to improve content visibility in generative-engine responses. However, prior studies mainly analyze final responses, leaving the internal execution process and its variation across model settings less understood.
To this end, we conduct an observational study of four ChatGPT Web settings from ChatGPT 5.2 to 5.5, covering Instant and Thinking modes at the URL, hostname, and entity levels. We examine how ChatGPT rewrites queries, retrieves candidate webpages, and selects references. The Instant settings remain mostly single-round, while GPT-5.4 Thinking frequently performs multiple retrieval rounds. Across repeated runs, candidates retrieved more often are also more likely to be cited.
We then evaluate Google Search and the OpenAI Responses API as external proxies. Responses API candidates provide broader coverage, while its citations more closely match sources cited by ChatGPT Web. Both proxies match ChatGPT Web better at the hostname and entity levels than at the exact-URL level. Even when combined, they recover only about one quarter of cited URLs. We further test whether these external signals remain predictive when combined. Within the external candidate set, a supervised model combining them reaches AUCs above 0.74 for GPT-5.3 and GPT-5.5 at all three levels.
These findings suggest that GEO should distinguish retrieval visibility from citation visibility, account for rewritten queries and run-to-run variation, and consider hostname- and entity-level presence in addition to exact URLs. Differences across settings also suggest that GEO strategies should be re-evaluated when the deployed setting or serving environment changes.
\end{abstract}

\maketitle
\section{Introduction}

Web search is widely used to access online information. In traditional search engines such as Google Search, webpage visibility mainly depends on retrieval and ranking. In contrast, generative engines such as ChatGPT Web introduce an extra citation selection step. They retrieve webpages to generate an answer, but only a subset of the retrieved webpages are shown to users as citations. Understanding how generative engines select webpages for citation, and which signals remain informative across different model settings, can help brands improve their visibility and share of voice in generative-engine responses, a practice known as \textit{generative engine optimization} (GEO)~\cite{aggarwal2024geo}.
Although prior studies have extensively measured generative engines, they mainly focus on final responses, such as which webpages are cited~\cite{zhang2026measurement}, how cited sources differ across search systems~\cite{kirsten2025characterizing,grossman2026disrupts}, and whether different systems rely on the same sources~\cite{yang2025news,huang2026answerbubbles}. However, these studies do not examine how a generative engine executes a user query or whether the observed process differs across model settings.

To this end, we conduct an observational study of four ChatGPT Web model settings from ChatGPT 5.2 to 5.5, spanning Instant and Thinking modes. We examine retrieval and citation behavior at the URL, hostname, and entity levels. We first examine how ChatGPT rewrites user queries, retrieves candidate webpages, and selects references from server-sent event (SSE) traces. We then evaluate whether Google Search and the OpenAI Responses API can approximate this process from externally observable results. We find that the observed model settings differ in how ChatGPT Web rewrites queries and retrieves webpages. Among the Instant settings, GPT-5.3 and GPT-5.5 use fewer executed queries than GPT-5.2 and remain mostly single-round, while the GPT-5.4 Thinking setting frequently performs multiple retrieval rounds and narrows the search target in later rounds. Repeated runs retrieve many different webpages, but candidates that appear more often across runs are also more likely to be cited. Longer retrieval snippets provide a weaker but consistent citation signal. We find that Google Search and the OpenAI Responses API only partially reproduce ChatGPT Web. Among the proxies, Responses API candidates provide broader coverage of sources cited by ChatGPT Web, while Responses API citations more closely match them. Both proxies match ChatGPT Web more closely at the hostname and entity levels than at the URL level.

Finally, we examine whether the external signals identified above remain predictive when considered together. We combine signals from Google Search and the OpenAI Responses API in a supervised model to test their joint predictive information. The resulting AUCs are 0.782, 0.763, and 0.743 on \webthree\ at the URL, hostname, and entity levels, and 0.779, 0.763, and 0.742 on \webfive. This combination achieves the highest AUC in five of the six setting-granularity combinations compared with rankings based on Google Search rank, Responses API rank, and Responses API citations. Rank, Responses API exposure frequency, citation conversion, multi-URL support, and snippet length account for most of the predictive importance. However, these results are conditional on external candidate coverage. Google Search and the OpenAI Responses API together recover only 28.3\% and 26.2\% of the cited URLs for \webthree\ and \webfive, compared with 50.3\% and 48.7\% of the cited hostnames.
These findings suggest that GEO should distinguish retrieval visibility from citation visibility, account for rewritten queries and run-to-run variation, and consider hostname- and entity-level presence in addition to exact URLs. Since retrieval behavior, proxy effectiveness, and citation signals differ across observed settings, GEO strategies should be re-evaluated when the deployed model setting or serving environment changes.

To summarize, we make the following contributions.
\begin{itemize}[topsep=2pt,itemsep=1pt,parsep=0pt,partopsep=0pt,leftmargin=11pt]
    \item We conduct an observational study of query rewriting, candidate retrieval, and citation selection across four ChatGPT Web model settings (Section~\ref{sec:pipeline}).
    \item We evaluate Google Search and the OpenAI Responses API as proxies at the URL, hostname, and entity levels (Section~\ref{sec:proxy}).
    \item We test whether external proxy signals jointly provide predictive information about ChatGPT Web citation (Section~\ref{sec:exp}).
    \item We discuss implications for GEO monitoring and optimization (Section~\ref{sec:geo}).
\end{itemize}

\section{Data Collection and Preprocessing}
\label{sec:prelim}

\stitle{Query set}
To ground our query set in real user behavior, we draw queries from Search Arena~\cite{miroyan2025search}, a crowd-sourced dataset of 24k in-the-wild user interactions with search-augmented LLMs. We sample 80 queries across four intent types and two languages, covering \textit{Factual Lookup}, \textit{Information Synthesis}, \textit{Analysis}, and \textit{Recommendation}. Each intent contains 20 queries, evenly split into 10 Chinese and 10 English queries. We exclude the remaining intents, as these either rely on parametric knowledge rather than retrieval, or target a destination rather than a cited answer~\cite{broder2002taxonomy,rose2004understanding}.

\stitle{ChatGPT Web}
We collect SSE traces under four ChatGPT Web model settings, i.e., \webtwo, \webthree, \webfour, and \webfive, to compare their observed retrieval and citation behavior. From each trace, we extract the executed queries, retrieved webpages, supporting webpages, and final citations. Between 2026-03-29 and 2026-07-29, \webthree\ and \webfive\ each cover the same 80 queries with 20 independent runs per query. In contrast, \webtwo\ and \webfour\ cover 32 and 30 queries with 7--28 and 4--20 runs per query, respectively, because they were deprecated before the remaining runs could be collected. Table~\ref{tab:dataset} summarizes the resulting dataset.

\stitle{Source sets and citations}
For each run, $W_{\mathrm{ret}}$ contains webpages returned as main search results or references, while $W_{\mathrm{sup}}$ contains supporting webpages explicitly attached to these references. We refer to the reference to which a supporting webpage is attached as its \textit{parent reference}. We define the candidate pool as $W_{\mathrm{pool}}=W_{\mathrm{ret}}\cup W_{\mathrm{sup}}$. We define $W_{\mathrm{cite}}$ as the selected reference URLs and the supporting webpages attached to them. When the final citation list is available, the selected reference URLs are those displayed directly in that list. When the list is absent for a completed stream, we use the completed grouped references as a fallback. Thus, $W_{\mathrm{cite}}$ follows the SSE structure of selected references and their supporting webpages rather than including only URLs displayed separately to users. A supporting webpage is therefore considered cited when its parent reference is selected, even if the supporting URL is not displayed separately. For each $u\in W_{\mathrm{pool}}$, we set the citation label $y_W(u)=1$ if $u\in W_{\mathrm{cite}}$ and set it to $0$ otherwise. Unless stated otherwise, subsequent analyses use $W_{\mathrm{pool}}$ as the source universe and $y_W$ as the citation label.

\begin{table}[t]
    \centering
    \small
    \setlength{\tabcolsep}{6pt}
    \caption{Dataset coverage by ChatGPT Web setting.}
    \vspace{-2mm}
    \label{tab:dataset}
    \begin{tabular}{lrrrr}
        \toprule
        Setting & \#intents & \#queries & \#repeats & \#runs \\
        \midrule
        \webtwo    & 4 & 32 & 7--28 & 547 \\
        \webthree  & 4 & 80 & 20 & 1600 \\
        \webfour   & 4 & 30 & 4--20 & 491 \\
        \webfive   & 4 & 80 & 20 & 1600 \\
        \bottomrule
    \end{tabular}
    \vspace{-4mm}
\end{table}

\stitle{OpenAI Responses API and Google Search}
A straightforward way to obtain a proxy for ChatGPT Web is to query comparable models through the OpenAI Responses API. Specifically, we invoke \texttt{\seqsplit{gpt-5.3-chat-latest}} and \texttt{\seqsplit{gpt-5.5}} with web search enabled and no system prompt. For \texttt{\seqsplit{gpt-5.5}}, we additionally set \texttt{\seqsplit{reasoning.effort="none"}} to make the inference configuration closer to the Instant setting used by ChatGPT Web. Following the per-query replication used for ChatGPT Web, we independently invoke each query 20 times for each model. From each response, we extract candidate webpages from \texttt{\seqsplit{web\_search\_call.action.sources[]}} and cited webpages from \texttt{\seqsplit{url\_citation}}. In total, we make 1{,}600 Responses API calls for each model. Each \texttt{\seqsplit{gpt-5.3-chat-latest}} response contains an average of 27.9 candidate entries and 5.8 URL-citation annotations, while each \texttt{\seqsplit{gpt-5.5}} response contains an average of 25.7 candidate entries and 5.4 URL-citation annotations.
As a second external proxy, we use Google Search. Specifically, we retrieve five pages of organic results per query through SerpAPI~\cite{serpapi2026}. The resulting Google Search snapshots contain an average of 43.2 organic results per query, as some result pages are empty. For each query, we collect the ChatGPT Web SSE traces and the results from both external proxies within the same three-day window.

\begin{figure*}[!t]
    \centering
    \vspace{-2mm}
    \includegraphics[width=\textwidth]{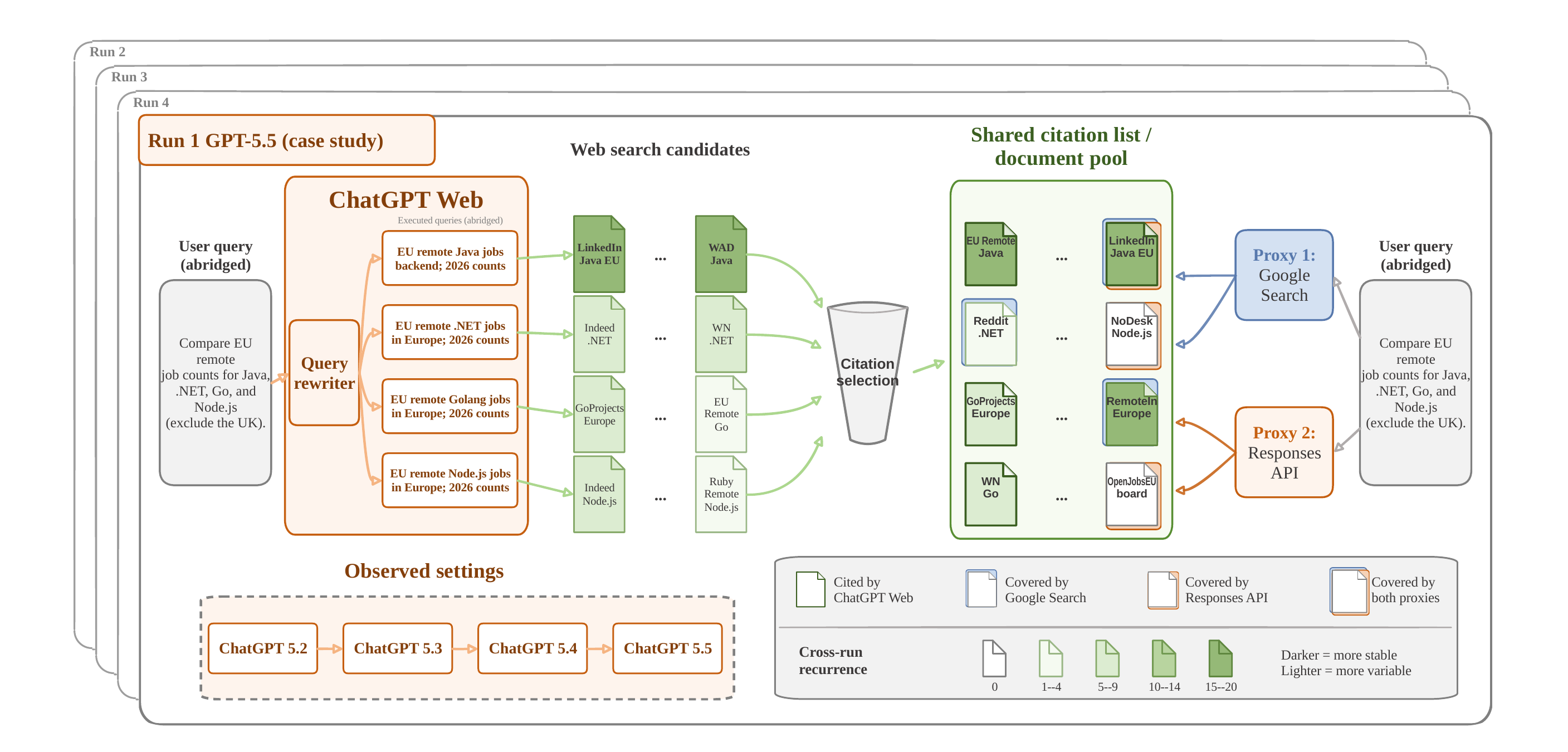}
    \vspace{-4mm}
    \caption{Overview of ChatGPT Web probing and proxy evaluation. The GPT-5.5 run is an abridged case study. Document darkness encodes cross-run occurrence. Blue marks Google Search coverage; orange marks Responses API coverage.}
    \vspace{-2mm}
    \label{fig:pipeline}
\end{figure*}

\stitle{Data normalization}
Across ChatGPT Web and the two external proxies, we normalize sources at the URL, hostname, and entity levels under a unified procedure. We refer to the URL-level procedure as \texttt{\seqsplit{deep\_normalize}}. It decodes each URL, lowercases its scheme and hostname, and removes a leading \texttt{www.}, default ports, trailing path slashes, fragments, and known tracking parameters such as \texttt{\seqsplit{utm\_*}}. For hostnames, we use the normalized hostname while retaining subdomains, so \texttt{blog.example.com} and \texttt{example.com} are treated as distinct hostnames.
At the entity level, we extract named entities from the title and snippet associated with each query--URL record. For English records, spaCy \texttt{\seqsplit{en\_core\_web\_sm}} identifies candidate spans, which are then filtered by \texttt{\seqsplit{GPT-4.1-mini}}. For Chinese records, \texttt{\seqsplit{GPT-4.1-mini}} directly extracts entities from the same context and maps Chinese names to lowercase English forms when possible. We then normalize entity mentions and deduplicate aliases and overlapping variants. Finally, we manually inspect the extracted entities and normalized aliases to verify the quality of the resulting entity annotations before aggregation.
Let $E_q(u)$ denote the resulting entity set for URL $u$ under query $q$. We do not infer entities from a URL's hostname when its title and snippet yield no entity. Brands and products are included as entity types rather than annotated separately. Within ChatGPT Web run $r$, we set $y_W^E(e,r)=1$ if at least one URL $u\in W_{\mathrm{pool}}$ satisfies $e\in E_q(u)$ and $y_W(u)=1$, and set it to $0$ for other entities appearing in the run.

\stitle{Statistical tests}
For matched comparisons between \webthree\ and \webfive, we use two-sided paired permutation tests at the query level and apply Holm correction across granularities within each metric. We regard adjusted $p<0.05$ as statistically significant.

\section{Execution Pipeline}\label{sec:pipeline}

\begin{table}[t]
    \centering
    \small
    \caption{{Pipeline statistics: the average number of executed queries (\#queries), main candidate URLs in \(W_{\mathrm{ret}}\) (\#main cand.), and cited URLs assigned \(y_W=1\) (\#cited).}}
    \vspace{-2mm}
    \label{tab:pipeline-stats}
    \begin{tabular}{lccc}
        \toprule
        Setting & \#queries & \#main cand. & {\#cited}
        \\
        \midrule
        \webtwo   & {2.36} & {39.38} & {5.91} \\
        \webthree & {1.06} & {18.74} & {5.55} \\
        \webfour  & {11.82} & {73.51} & {7.81} \\
        \webfive  & {1.45} & {23.47} & {6.53} \\
        \bottomrule
    \end{tabular}
    \vspace{-2mm}
\end{table}

\subsection{Pipeline Overview}
\label{sec:pipeline-overview}

From the SSE traces, we observe three components in the execution of a ChatGPT Web run, as illustrated in Figure~\ref{fig:pipeline}. First, \textit{query rewriting} reformulates the user query into one or more executed queries that determine what is searched. Second, web retrieval returns main candidate webpages in $W_{\mathrm{ret}}$, while grouped references may attach supporting webpages in $W_{\mathrm{sup}}$. Together, they form the observed candidate pool $W_{\mathrm{pool}}$. Third, answer synthesis selects references for the final answer, recorded together with their supporting webpages in $W_{\mathrm{cite}}$. We refer to the mapping from retrieved candidates to cited sources as the \textit{candidate-to-citation funnel}. We analyze this funnel using the citation label $y_W$ defined in Section~\ref{sec:prelim}. Query rewriting and retrieval may also interleave within a response, \ie an additional executed query may be issued after an initial retrieval round. Thus, the three components in Figure~\ref{fig:pipeline} represent an analytical decomposition rather than a strictly linear sequence.

Section~\ref{sec:query-rewriting} examines query rewriting, while Section~\ref{sec:funnel} analyzes candidate retrieval and citation selection together. Table~\ref{tab:pipeline-stats} summarizes the stage-level statistics. Code and data are available at {\color{blue}\url{https://anonymous.4open.science/r/chatgpt-citation-prediction-E45F/}}.

\subsection{Query Rewriting}
\label{sec:query-rewriting}

We distinguish three units in query rewriting.
A \textit{run} is one complete ChatGPT Web response, a \textit{retrieval round} is one use of the web search tool, and an \textit{executed query} is one query issued within a retrieval round.
Since one round may issue multiple queries, Table~\ref{tab:query-rewriting-overview} reports queries and rounds separately.
Executed queries are observable in only 1{,}339 of the 1{,}600 \webthree\ runs because the backend queries are unavailable for the remaining runs.

\stitle{The observed Instant settings remain mostly single-round but differ in query count and language}
As shown in Table~\ref{tab:query-rewriting-overview}, \webtwo\ issues multiple queries in every observed run, compared with 6.2\% for \webthree\ and 31.3\% for \webfive.
All three Instant settings remain mostly single-round, with multi-round rates of 17.4\%, 5.5\%, and 3.8\%, respectively.
Thus, \webfive\ issues multiple queries more often than \webthree\ without using more retrieval rounds.
Query language also differs across the observed Instant settings.
For Chinese questions, 96.8\% of \webtwo\ runs issue both Chinese and English queries, whereas 67.0\% and 69.6\% of \webthree\ and \webfive\ runs use English-only queries.
An exact match to the original user query occurs in 5.5\% of \webtwo\ runs and in none of the observed \webthree\ or \webfive\ runs.

\begin{table}[t]
    \centering
    \small
    \setlength{\tabcolsep}{5pt}
    \caption{Query rewriting across ChatGPT Web settings.}
    \label{tab:query-rewriting-overview}
    \vspace{-2mm}
    \begin{tabular}{lrrrr}
        \toprule
        & 5.2-instant & 5.3-instant & 5.4-thinking & 5.5-instant \\
        \midrule
        Observed/All   & 547/547 & 1339/1600 & 491/491 & 1600/1600 \\
        Queries/run    & 2.36 & 1.06 & 11.82 & 1.45 \\
        Multi-query    & 100.0\% & 6.2\% & 99.4\% & 31.3\% \\
        Rounds/run     & 1.18 & 1.06 & 3.20 & 1.04 \\
        Multi-round    & 17.4\% & 5.5\% & 71.3\% & 3.8\% \\
        Exact match    & 5.5\% & 0.0\% & 0.0\% & 0.0\% \\
        \bottomrule
    \end{tabular}
    \vspace{-2mm}
\end{table}

\stitle{The GPT-5.4 Thinking setting uses multi-round retrieval and narrows the search target}
We treat \webfour\ as a supplementary setting because its 30-query set is not matched to the Instant-setting query sets.
Across 491 runs, \webfour\ issues 11.82 queries over 3.20 retrieval rounds per run on average, and 71.3\% of its runs use the web search tool more than once.
Multi-round retrieval occurs in 87.6\% of Information Synthesis runs, 85.2\% of Analysis runs, and 74.4\% of Recommendation runs, compared with 10.8\% of Factual Lookup runs.
These intent-level results are descriptive because the groups contain different numbers of queries and runs.
Later rounds increasingly narrow the retrieval target.
Among the 4{,}077 executed queries after the first round, only 13 (0.3\%) repeat an earlier query.
The share containing \texttt{\seqsplit{site:}} increases from 25.1\% in the first round to 54.7\% in later rounds, while quoted phrases increase from 1.0\% to 22.7\%.
These results show that later rounds in \webfour\ use more targeted query forms, but do not establish that additional queries or rounds improve candidate quality or citation selection.

\subsection{Candidate-to-Citation Funnel}
\label{sec:funnel}

We next examine which sources in $W_{\mathrm{pool}}$ receive the citation label $y_W=1$ at the URL, hostname, and entity levels.
For hostnames, the label is positive if any URL on the hostname has $y_W=1$, while entities use $y_W^E$ as defined in Section~\ref{sec:prelim}.
Unless stated otherwise, the analyses below use the 80 aligned queries for \webthree\ and \webfive, with 20 runs per query and query-macro estimates.

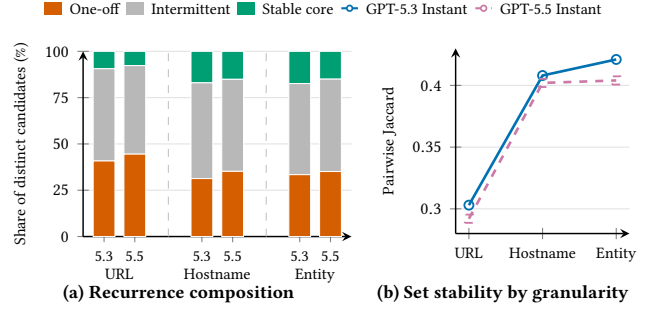
\begin{figure}[t]
    \centering
    \begingroup
    \definecolor{stabilityOneOff}{HTML}{D55E00}
    \definecolor{stabilityIntermittent}{HTML}{B8B8B8}
    \definecolor{stabilityCore}{HTML}{009E73}
    \definecolor{stabilityFiveThree}{HTML}{0072B2}
    \definecolor{stabilityFiveFive}{HTML}{CC79A7}

    \newcommand{\stabilityswatch}[2]{%
        \tikz[baseline=-0.55ex]{\node[fill=#1, minimum width=6.5pt, minimum height=4.6pt, inner sep=0pt] {};}
        \hspace{0.08em}{\scriptsize #2}%
    }
    \newcommand{\stabilityline}[3]{%
        \tikz[baseline=-0.55ex]{
            \draw[#1, line width=0.8pt, #3] (0,0) -- (0.22,0);
            \filldraw[fill=white, draw=#1, line width=0.65pt] (0.11,0) circle[radius=1.05pt];
        }
        \hspace{0.05em}{\scriptsize #2}%
    }

    \makebox[\columnwidth][c]{%
        \stabilityswatch{stabilityOneOff}{One-off}\hspace{0.28em}%
        \stabilityswatch{stabilityIntermittent}{Intermittent}\hspace{0.28em}%
        \stabilityswatch{stabilityCore}{Stable core}\hspace{0.28em}%
        \stabilityline{stabilityFiveThree}{GPT-5.3 Instant}{solid}\hspace{0.28em}%
        \stabilityline{stabilityFiveFive}{GPT-5.5 Instant}{dashed}%
    }\par\vspace{-0.35em}

    \begin{minipage}[t]{0.55\columnwidth}
        \vspace{0pt}
        \centering
        \begin{tikzpicture}
            \begin{axis}[
                width=3.55cm,
                height=2.45cm,
                scale only axis,
                ybar stacked,
                bar width=8.2pt,
                xmin=-0.52,
                xmax=3.32,
                ymin=0,
                ymax=100,
                ytick={0,25,50,75,100},
                ylabel={Share of distinct candidates (\%)},
                ylabel style={font=\scriptsize},
                xtick={-0.22,0.22,1.18,1.62,2.58,3.02},
                xticklabels={5.3,5.5,5.3,5.5,5.3,5.5},
                xticklabel style={font=\scriptsize},
                extra x ticks={0,1.4,2.8},
                extra x tick labels={URL,Hostname,Entity},
                extra x tick style={
                    tick style={draw=none},
                    tick label style={font=\scriptsize\bfseries, yshift=-1.75ex},
                },
                tick label style={font=\scriptsize},
                tick style={line width=0.4pt},
                axis lines=left,
                axis line style={line width=0.55pt},
                ymajorgrids=true,
                grid style={gray!24, line width=0.35pt},
                clip=false,
            ]
                \addplot[fill=stabilityOneOff, draw=white, line width=0.3pt] coordinates {
                    (-0.22,40.8) (0.22,44.5)
                    (1.18,31.3) (1.62,35.2)
                    (2.58,33.4) (3.02,35.1)
                };
                \addplot[fill=stabilityIntermittent, draw=white, line width=0.3pt] coordinates {
                    (-0.22,49.8) (0.22,47.8)
                    (1.18,51.7) (1.62,49.7)
                    (2.58,49.2) (3.02,49.9)
                };
                \addplot[fill=stabilityCore, draw=white, line width=0.3pt] coordinates {
                    (-0.22,9.4) (0.22,7.7)
                    (1.18,17.0) (1.62,15.1)
                    (2.58,17.4) (3.02,15.0)
                };
                \draw[gray!48, dashed, line width=0.4pt] (axis cs:0.70,0) -- (axis cs:0.70,100);
                \draw[gray!48, dashed, line width=0.4pt] (axis cs:2.10,0) -- (axis cs:2.10,100);
                \node[font=\tiny\bfseries, text=white] at (axis cs:-0.22,20.4) {40.8};
                \node[font=\tiny\bfseries, text=white] at (axis cs:0.22,22.25) {44.5};
                \node[font=\tiny\bfseries, text=white] at (axis cs:1.18,15.65) {31.3};
                \node[font=\tiny\bfseries, text=white] at (axis cs:1.62,17.6) {35.2};
                \node[font=\tiny\bfseries, text=white] at (axis cs:2.58,16.7) {33.4};
                \node[font=\tiny\bfseries, text=white] at (axis cs:3.02,17.55) {35.1};
                \node[font=\tiny\bfseries, text=white] at (axis cs:-0.22,95.3) {9.4};
                \node[font=\tiny\bfseries, text=white] at (axis cs:0.22,96.15) {7.7};
                \node[font=\tiny\bfseries, text=white] at (axis cs:1.18,91.5) {17.0};
                \node[font=\tiny\bfseries, text=white] at (axis cs:1.62,92.45) {15.1};
                \node[font=\tiny\bfseries, text=white] at (axis cs:2.58,91.3) {17.4};
                \node[font=\tiny\bfseries, text=white] at (axis cs:3.02,92.5) {15.0};
            \end{axis}
        \end{tikzpicture}
    \end{minipage}\hfill
    \begin{minipage}[t]{0.42\columnwidth}
        \vspace{0.17cm}
        \centering
        \begin{tikzpicture}
            \begin{axis}[
                width=2.30cm,
                height=2.45cm,
                scale only axis,
                xmin=-0.18,
                xmax=2.18,
                ymin=0.28,
                ymax=0.43,
                xtick={0,1,2},
                xticklabels={URL,Hostname,Entity},
                ytick={0.30,0.35,0.40},
                ylabel={Pairwise Jaccard},
                ylabel style={font=\scriptsize},
                xticklabel style={font=\scriptsize},
                tick label style={font=\scriptsize},
                tick style={line width=0.4pt},
                axis lines=left,
                axis line style={line width=0.55pt},
                ymajorgrids=true,
                grid style={gray!24, line width=0.35pt},
            ]
                \addplot[
                    color=stabilityFiveThree,
                    line width=1.0pt,
                    mark=o,
                    mark size=1.7pt,
                    mark options={fill=white, line width=0.7pt},
                ] coordinates {(0,0.303) (1,0.408) (2,0.421)};
                \addplot[
                    color=stabilityFiveFive,
                    dashed,
                    line width=1.0pt,
                    mark=square,
                    mark size=1.55pt,
                    mark options={fill=white, line width=0.7pt},
                ] coordinates {(0,0.292) (1,0.402) (2,0.404)};
            \end{axis}
        \end{tikzpicture}
    \end{minipage}
    \par\vspace{-0.35em}
    \begin{minipage}[t]{0.55\columnwidth}
        \centering
        \footnotesize\bfseries (a) Recurrence composition
    \end{minipage}\hfill
    \begin{minipage}[t]{0.42\columnwidth}
        \centering
        \footnotesize\bfseries (b) Set stability by granularity
    \end{minipage}
    \endgroup
    \vspace{-2mm}
    \caption{Cross-run recurrence and stability of sources in $W_{\mathrm{pool}}$. Panels (a) and (b) report recurrence composition and query-macro pairwise Jaccard similarity across runs, respectively.}
    \vspace{-2mm}
    \label{fig:candidate-stability}
\end{figure}

\stitle{Candidates vary across runs but are more stable at the hostname and entity levels}
We first measure how consistently query-specific candidates reappear across the 20 runs.
We refer to an item as a one-off if it appears in only one run and as part of the stable core if it appears in at least 14 runs.
The remaining items are classified as intermittent.
We also compute pairwise Jaccard similarity between runs to measure the overlap between their candidate sets.
All items are matched using the URL, hostname, and entity normalization defined in Section~\ref{sec:prelim}, and all statistics are computed per query and then macro-averaged.
As shown in Figure~\ref{fig:candidate-stability}, 40.8\% of \webthree\ URL candidates and 44.5\% of \webfive\ URL candidates appear in only one run, while 9.4\% and 7.7\% belong to the stable core.
Across the 20 runs, \webfive\ retrieves 132.2 distinct URLs per query on average, compared with 96.4 for \webthree, but URL-level Jaccard similarity is similar across the two settings, at 0.303 for \webthree\ and 0.292 for \webfive.
Thus, \webfive\ retrieves more distinct URLs without increasing exact-URL stability.
Candidate sets are more stable at coarser granularities.
Hostname-level Jaccard similarity increases to 0.408 for \webthree\ and 0.402 for \webfive, while entity-level similarity reaches 0.421 and 0.404, respectively.
This suggests that repeated runs often retrieve different pages with the same hostnames or entities rather than the same exact URLs.

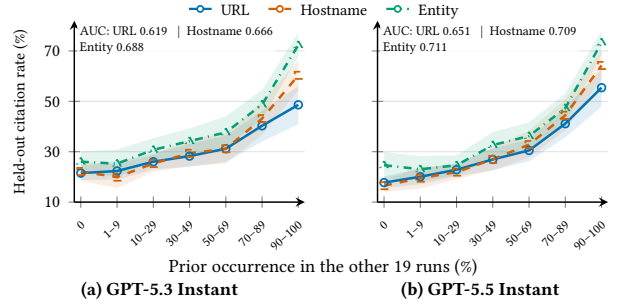
\begin{figure}[t]
    \centering
    \begingroup
    \definecolor{recurrenceURL}{HTML}{0072B2}
    \definecolor{recurrenceHostname}{HTML}{D55E00}
    \definecolor{recurrenceEntity}{HTML}{009E73}

    \newcommand{\recurrencelegend}[4]{%
        \tikz[baseline=-0.55ex]{
            \draw[#1, line width=0.85pt, #3] (0,0) -- (0.30,0);
            \filldraw[fill=white, draw=#1, line width=0.65pt] (0.15,0) circle[radius=1.15pt];
        }
        \hspace{0.1em}{\scriptsize #2}%
    }

    \begin{tabular}{@{}l@{\hspace{0.75em}}l@{\hspace{0.75em}}l@{}}
        \recurrencelegend{recurrenceURL}{URL}{solid}{circle} &
        \recurrencelegend{recurrenceHostname}{Hostname}{dashed}{rectangle} &
        \recurrencelegend{recurrenceEntity}{Entity}{dash dot}{regular polygon, regular polygon sides=3}
    \end{tabular}\par\vspace{-0.25em}

    \begin{minipage}[t]{0.49\columnwidth}
        \centering
        \begin{tikzpicture}
            \begin{axis}[
                width=3.05cm,
                height=2.4cm,
                scale only axis,
                xmin=-0.18,
                xmax=6.18,
                ymin=10,
                ymax=82,
                xtick={0,1,2,3,4,5,6},
                xticklabels={0,1--9,10--29,30--49,50--69,70--89,90--100},
                ytick={10,30,50,70},
                ylabel={{Held-out citation rate (\%)}},
                ylabel style={font=\scriptsize},
                xticklabel style={font=\tiny, rotate=42, anchor=north east},
                tick label style={font=\scriptsize},
                tick style={line width=0.4pt},
                axis lines=left,
                axis line style={line width=0.55pt},
                ymajorgrids=true,
                grid style={gray!24, line width=0.35pt},
                clip=true,
            ]
                \addplot[name path=url53upper, draw=none, forget plot] coordinates {
                    (0,24.35) (1,26.22) (2,29.18) (3,33.16) (4,36.44) (5,46.29) (6,56.09)
                };
                \addplot[name path=url53lower, draw=none, forget plot] coordinates {
                    (0,18.77) (1,18.88) (2,23.02) (3,23.98) (4,25.95) (5,34.02) (6,40.87)
                };
                \addplot[fill=recurrenceURL, fill opacity=0.12, draw=none, forget plot] fill between[of=url53upper and url53lower];

                \addplot[name path=host53upper, draw=none, forget plot] coordinates {
                    (0,26.16) (1,25.01) (2,28.84) (3,34.98) (4,37.33) (5,49.88) (6,66.39)
                };
                \addplot[name path=host53lower, draw=none, forget plot] coordinates {
                    (0,18.50) (1,15.60) (2,21.99) (3,24.40) (4,25.35) (5,36.76) (6,54.37)
                };
                \addplot[fill=recurrenceHostname, fill opacity=0.11, draw=none, forget plot] fill between[of=host53upper and host53lower];

                \addplot[name path=entity53upper, draw=none, forget plot] coordinates {
                    (0,30.10) (1,30.88) (2,35.43) (3,39.28) (4,44.07) (5,54.59) (6,77.33)
                };
                \addplot[name path=entity53lower, draw=none, forget plot] coordinates {
                    (0,21.89) (1,20.05) (2,26.23) (3,28.71) (4,31.65) (5,42.90) (6,67.70)
                };
                \addplot[fill=recurrenceEntity, fill opacity=0.11, draw=none, forget plot] fill between[of=entity53upper and entity53lower];

                \addplot[color=recurrenceURL, line width=0.95pt, mark=o, mark size=1.45pt, mark options={fill=white, line width=0.65pt}] coordinates {
                    (0,21.52) (1,22.34) (2,26.04) (3,28.27) (4,31.19) (5,40.31) (6,48.62)
                };
                \addplot[color=recurrenceHostname, dashed, line width=0.95pt, mark=square, mark size=1.35pt, mark options={fill=white, line width=0.65pt}] coordinates {
                    (0,22.16) (1,19.90) (2,25.33) (3,29.38) (4,31.17) (5,43.13) (6,60.33)
                };
                \addplot[color=recurrenceEntity, dash dot, line width=0.95pt, mark=triangle, mark size=1.55pt, mark options={fill=white, line width=0.65pt}] coordinates {
                    (0,26.04) (1,25.25) (2,30.75) (3,34.20) (4,37.68) (5,48.72) (6,72.31)
                };
                \node[anchor=north west, align=left, font=\tiny, fill=white, fill opacity=0.78, text opacity=1, inner sep=1pt] at (rel axis cs:0.01,0.99) {AUC: URL 0.619 \enspace|\enspace Hostname 0.666\\Entity 0.688};
            \end{axis}
        \end{tikzpicture}
    \end{minipage}\hfill
    \begin{minipage}[t]{0.49\columnwidth}
        \centering
        \begin{tikzpicture}
            \begin{axis}[
                width=3.05cm,
                height=2.4cm,
                scale only axis,
                xmin=-0.18,
                xmax=6.18,
                ymin=10,
                ymax=82,
                xtick={0,1,2,3,4,5,6},
                xticklabels={0,1--9,10--29,30--49,50--69,70--89,90--100},
                ytick={10,30,50,70},
                yticklabels={},
                xticklabel style={font=\tiny, rotate=42, anchor=north east},
                tick label style={font=\scriptsize},
                tick style={line width=0.4pt},
                axis lines=left,
                axis line style={line width=0.55pt},
                ymajorgrids=true,
                grid style={gray!24, line width=0.35pt},
                clip=true,
            ]
                \addplot[name path=url55upper, draw=none, forget plot] coordinates {
                    (0,20.19) (1,23.27) (2,25.55) (3,31.25) (4,34.96) (5,47.34) (6,62.88)
                };
                \addplot[name path=url55lower, draw=none, forget plot] coordinates {
                    (0,15.50) (1,17.09) (2,20.19) (3,22.66) (4,26.20) (5,35.71) (6,47.85)
                };
                \addplot[fill=recurrenceURL, fill opacity=0.12, draw=none, forget plot] fill between[of=url55upper and url55lower];

                \addplot[name path=host55upper, draw=none, forget plot] coordinates {
                    (0,19.05) (1,23.84) (2,25.00) (3,31.65) (4,38.37) (5,50.88) (6,70.11)
                };
                \addplot[name path=host55lower, draw=none, forget plot] coordinates {
                    (0,14.12) (1,15.29) (2,19.10) (3,22.62) (4,27.69) (5,38.50) (6,58.75)
                };
                \addplot[fill=recurrenceHostname, fill opacity=0.11, draw=none, forget plot] fill between[of=host55upper and host55lower];

                \addplot[name path=entity55upper, draw=none, forget plot] coordinates {
                    (0,29.88) (1,28.25) (2,28.52) (3,37.79) (4,41.68) (5,53.18) (6,77.95)
                };
                \addplot[name path=entity55lower, draw=none, forget plot] coordinates {
                    (0,19.63) (1,18.28) (2,20.67) (3,27.35) (4,30.89) (5,41.35) (6,68.52)
                };
                \addplot[fill=recurrenceEntity, fill opacity=0.11, draw=none, forget plot] fill between[of=entity55upper and entity55lower];

                \addplot[color=recurrenceURL, line width=0.95pt, mark=o, mark size=1.45pt, mark options={fill=white, line width=0.65pt}] coordinates {
                    (0,17.74) (1,20.08) (2,22.86) (3,26.94) (4,30.52) (5,41.14) (6,55.42)
                };
                \addplot[color=recurrenceHostname, dashed, line width=0.95pt, mark=square, mark size=1.35pt, mark options={fill=white, line width=0.65pt}] coordinates {
                    (0,16.52) (1,19.52) (2,21.91) (3,26.98) (4,32.84) (5,44.48) (6,64.23)
                };
                \addplot[color=recurrenceEntity, dash dot, line width=0.95pt, mark=triangle, mark size=1.55pt, mark options={fill=white, line width=0.65pt}] coordinates {
                    (0,24.65) (1,23.04) (2,24.61) (3,32.64) (4,36.23) (5,47.31) (6,73.37)
                };
                \node[anchor=north west, align=left, font=\tiny, fill=white, fill opacity=0.78, text opacity=1, inner sep=1pt] at (rel axis cs:0.01,0.99) {AUC: URL 0.651 \enspace|\enspace Hostname 0.709\\Entity 0.711};
            \end{axis}
        \end{tikzpicture}
    \end{minipage}
    \par\vspace{-0.85em}
    \makebox[\columnwidth][c]{\footnotesize Prior occurrence in the other 19 runs (\%)}\par
    \vspace{-0.25em}
    \begin{minipage}[t]{0.49\columnwidth}
        \centering
        \footnotesize\bfseries (a) GPT-5.3 Instant
    \end{minipage}\hfill
    \begin{minipage}[t]{0.49\columnwidth}
        \centering
        \footnotesize\bfseries (b) GPT-5.5 Instant
    \end{minipage}
    \endgroup
    \vspace{-2mm}
    \caption{Held-out citation rate by leave-one-run-out recurrence. Shaded regions show 95\% query-bootstrap confidence intervals.}
    \vspace{-2mm}
    \label{fig:recurrence-citation}
\end{figure}

\stitle{Candidates retrieved more often across runs are more likely to be cited}
A within-sample analysis would give frequently retrieved items more opportunities to receive a positive label.
To avoid this overlap, we use a leave-one-run-out design.
For each item appearing in a held-out run, we measure how often it appears in the other 19 runs and use this recurrence frequency to predict its label in the held-out run.
The label is $y_W$ for URLs, is positive for a hostname if any URL on that hostname has $y_W=1$, and is $y_W^E$ for entities.
We compute AUC separately for each query and obtain 95\% confidence intervals from 2,000 query-level bootstrap samples.
As shown in Figure~\ref{fig:recurrence-citation}, cross-run recurrence predicts citation in the held-out run for both \webthree\ and \webfive.
The URL, hostname, and entity AUCs are 0.619, 0.666, and 0.688 for \webthree, and 0.651, 0.709, and 0.711 for \webfive, respectively.
The AUC increases from \webthree\ to \webfive\ at the URL and hostname levels remain significant after Holm correction (adjusted $p\leq0.026$).
Recurrence is therefore more informative at the hostname and entity levels than at the exact-URL level.
This is consistent with the stability results above, where repeated runs often retrieve different pages from the same hostnames or associated with the same entities.
The entity-level AUC exceeds the hostname-level AUC by only 0.021 for \webthree\ and 0.003 for \webfive.
The positive relationship also holds across all four query intents, although the highest-AUC intent differs between the two settings.
We therefore do not infer a consistent ordering across intents.
Finally, recurrence is observational and does not imply that a source would be cited if it appeared more often across repeated runs.

\stitle{Cited candidates tend to have longer retrieval snippets}
We compare the character lengths of cited and uncited nonempty snippets within the same run, yielding 28,368 candidate instances for \webthree\ and 33,976 for \webfive.
Using snippet length as the only score, we compute AUC separately for each query and obtain 95\% confidence intervals from 2,000 query-level bootstrap samples.
Snippet length yields AUCs of 0.612 for \webthree\ and 0.625 for \webfive, weaker than the cross-run recurrence signal above.
The difference is most visible in the right tail.
In \webthree\ and \webfive, 16.5\% and 16.9\% of cited snippets contain at least 500 characters, compared with 1.5\% and 1.9\% of uncited snippets.
The association remains positive when restricting the analysis to main candidates in $W_{\mathrm{ret}}$, with AUCs of 0.602 and 0.607.
Thus, the pattern is not driven solely by cited supporting pages.
Longer snippets may provide self-contained evidence for answer synthesis, but length does not measure semantic completeness and may reflect source type, structure, or retrieval position.
We therefore treat snippet length as an observable selection signal rather than a causal factor.

\stitle{Other signals are weaker or less consistent across settings}
Among the remaining attributes, retrieval position shows the clearest citation signal, with within-run AUC increasing from 0.561 for \webthree\ to 0.623 for \webfive.
However, candidates beyond position 15 still have citation rates of 24.0\% and 19.0\%, respectively, showing that position alone does not determine citation.
Publication recency is weaker, with AUCs of 0.482 and 0.535 and both 95\% confidence intervals including 0.5.
{Among reference types present in both settings (search, news, and academia), the share of distinct query--URL candidates cited in at least one run is highest for news in \webthree\ (52.0\%) and academia in \webfive\ (49.2\%).}
Factual Lookup has the highest cross-run Jaccard similarity of cited source sets in both settings, but the remaining intents show no consistent ordering.
We therefore treat these attributes as supplementary rather than general citation signals.


\section{Proxy Evaluation}
\label{sec:proxy}
\subsection{Setup and Metrics}

We evaluate three proxy sets.
$G$ contains up to five pages of organic Google Search results returned through SerpAPI. $A_{\mathrm{cand}}$ contains all sources exposed by the OpenAI Responses API, including citation-only URLs, while $A_{\mathrm{cite}}$ contains only its final citations. We additionally evaluate $G\cup A_{\mathrm{cand}}$ as a coverage-oriented union. Since the union has no predefined ranking, we evaluate it only as an unranked set.

For \webthree\ and \webfive, we use the same 80 queries, with 20 ChatGPT Web runs and 20 Responses API runs per query. At the URL and hostname levels, an item is relevant if it receives $y_W=1$ in at least one ChatGPT Web run. Its graded relevance is the number of ChatGPT Web runs in which it receives this label. We rank $G$ by its native Google Search order, $A_{\mathrm{cand}}$ by cross-run exposure frequency, and $A_{\mathrm{cite}}$ by cross-run citation frequency. URLs are matched after applying \texttt{\seqsplit{deep\_normalize}} as defined in Section~\ref{sec:prelim}, while hostnames retain subdomains and are deduplicated on both sides.
At the entity level, we aggregate evidence from the URL-to-entity mapping $E_q(u)$ defined in Section~\ref{sec:prelim}. For system $S\in\{W,A\}$, where $W$ denotes ChatGPT Web and $A$ denotes the OpenAI Responses API, let $a_S(u)$ denote the number of runs in which URL $u$ is exposed. We let $c_W(u)$ count ChatGPT Web runs in which $y_W(u)=1$, and let $c_A(u)$ count Responses API runs in which $u$ appears as a direct API citation. We sum these counts over all URLs associated with entity $e$:
$$ X_S(e)=\sum_{u:e\in E_q(u)}a_S(u) \quad\text{and}\quad C_S(e)=\sum_{u:e\in E_q(u)}c_S(u). $$
Here, $X_S(e)$ measures entity exposure. $C_W(e)$ aggregates support from cited URLs in ChatGPT Web, while $C_A(e)$ aggregates support from direct Responses API citations. An entity is relevant if $C_W(e)>0$, and its graded relevance is $C_W(e)$. We rank $G$ entities by exposure and native Google Search position, $A_{\mathrm{cand}}$ entities by $X_A(e)$, and $A_{\mathrm{cite}}$ entities by $C_A(e)$.

We report full-set Precision, Recall, and F1 to measure whether a proxy contains the sources cited by ChatGPT Web. We use linear-gain NDCG@$k$ to measure whether it ranks sources cited more frequently by ChatGPT Web near the top, where $k\in\{1,3,5,10\}$. All metrics are computed per query and then macro-averaged.

\subsection{URL and Hostname Levels}

\stitle{{Responses API candidates lead in recall, while citations lead in precision}}
Table~\ref{tab:channel-proxy-overall} shows a consistent trade-off between coverage and precision. {Among individual proxies,} $A_{\mathrm{cand}}$ achieves the highest Recall, while $A_{\mathrm{cite}}$ achieves the highest Precision and F1. For ranking, $A_{\mathrm{cite}}$ generally achieves higher NDCG than $A_{\mathrm{cand}}$, indicating that its ranking more closely follows ChatGPT Web citation frequency. The union $G\cup A_{\mathrm{cand}}$ provides the broadest coverage, but its lower Precision prevents a corresponding F1 improvement. Google Search adds coverage to the union but is generally weaker as a standalone proxy. Thus, the Responses API provides different signals at the candidate and citation stages.

\stitle{External proxies recover cited hostnames more reliably than exact URLs}
Proxy effectiveness is consistently higher at the hostname level than at the exact-URL level. For example, the F1 of $A_{\mathrm{cand}}$ increases from 9.5\% to 19.9\% for \webthree\ and from 10.3\% to 22.4\% for \webfive, with the same general pattern in NDCG. The difference is also clear in coverage. The union $G\cup A_{\mathrm{cand}}$ recovers 28.3\% and 26.2\% of cited URLs for \webthree\ and \webfive, respectively, compared with 50.3\% and 48.7\% of cited hostnames. External proxies therefore recover the same hostnames more reliably than the same exact URLs.
\begin{table}[t]
\centering
\small
\setlength{\tabcolsep}{2.1pt}
\renewcommand{\arraystretch}{1.06}
\caption{Proxy effectiveness at the URL and hostname levels.
Panels (a) and (b) report full-set visibility and NDCG ranking quality, respectively.}
\label{tab:channel-proxy-overall}
\vspace{-2mm}

\textbf{(a) Full-set visibility}\\[2pt]
\begin{tabular}{@{\extracolsep{\fill}}llrrrrrr@{}}
    \toprule
    & & \multicolumn{3}{c}{URL} & \multicolumn{3}{c}{Hostname} \\
    \cmidrule(lr){3-5}
    \cmidrule(lr){6-8}
    Setting & Proxy
    & Precision & Recall & F1
    & Precision & Recall & F1 \\
    \midrule
    5.3 & $G$
        & 6.5 & 9.4 & 7.3
        & 17.1 & 26.1 & 19.2 \\
    5.3 & $A_{\mathrm{cand}}$
        & 6.6 & 22.9 & 9.5
        & 14.8 & 38.3 & 19.9 \\
    5.3 & $A_{\mathrm{cite}}$
        & 16.4 & 12.3 & 13.1
        & 27.7 & 23.3 & 23.3 \\
    5.3 & $G\cup A_{\mathrm{cand}}$
        & 5.9 & 28.3 & 9.3
        & 13.3 & 50.3 & 19.9 \\
    \midrule
    5.5 & $G$
        & 4.1 & 6.4 & 4.6
        & 12.5 & 22.8 & 14.8 \\
    5.5 & $A_{\mathrm{cand}}$
        & 7.1 & 22.5 & 10.3
        & 17.1 & 38.1 & 22.4 \\
    5.5 & $A_{\mathrm{cite}}$
        & 23.3 & 13.0 & 15.2
        & 40.5 & 23.9 & 27.7 \\
    5.5 & $G\cup A_{\mathrm{cand}}$
        & 5.7 & 26.2 & 9.0
        & 12.8 & 48.7 & 19.5 \\
    \bottomrule
\end{tabular}

\vspace{6pt}
\textbf{(b) Ranking quality (NDCG)}\\[2pt]
\begin{tabular}{@{\extracolsep{\fill}}llrrrrrrrr@{}}
    \toprule
    & & \multicolumn{4}{c}{URL} & \multicolumn{4}{c}{Hostname} \\
    \cmidrule(lr){3-6}
    \cmidrule(lr){7-10}
    Setting & Proxy
    & @1 & @3 & @5 & @10
    & @1 & @3 & @5 & @10 \\
    \midrule
    5.3 & $G$
        & 0.08 & 0.08 & 0.09 & 0.09
        & 0.19 & 0.18 & 0.20 & 0.21 \\
    5.3 & $A_{\mathrm{cand}}$
        & 0.12 & 0.13 & 0.14 & 0.15
        & 0.40 & 0.35 & 0.35 & 0.36 \\
    5.3 & $A_{\mathrm{cite}}$
        & 0.22 & 0.20 & 0.20 & 0.20
        & 0.42 & 0.37 & 0.35 & 0.35 \\
    \midrule
    5.5 & $G$
        & 0.03 & 0.05 & 0.05 & 0.05
        & 0.15 & 0.14 & 0.14 & 0.15 \\
    5.5 & $A_{\mathrm{cand}}$
        & 0.22 & 0.18 & 0.17 & 0.17
        & 0.42 & 0.40 & 0.39 & 0.38 \\
    5.5 & $A_{\mathrm{cite}}$
        & 0.27 & 0.26 & 0.25 & 0.25
        & 0.50 & 0.47 & 0.44 & 0.42 \\
    \bottomrule
\end{tabular}
\vspace{-2mm}
\end{table}

\stitle{{Responses API citations align more closely with \webfive}}
The largest difference between the \webthree\ and \webfive\ settings occurs for $A_{\mathrm{cite}}$ in both visibility and ranking. From \webthree\ to \webfive, URL Precision increases from 16.4\% to 23.3\%, while hostname Precision increases from 27.7\% to 40.5\%. Both increases remain significant after Holm correction (adjusted $p\leq0.026$). NDCG@10 also increases from 0.203 to 0.246 at the URL level and from 0.352 to 0.424 at the hostname level. In contrast, the Recall of $A_{\mathrm{cand}}$ remains nearly unchanged at both granularities. Thus, $A_{\mathrm{cite}}$ better matches both the sources cited by \webfive\ and their relative citation frequency, without a corresponding increase in candidate coverage.

\subsection{Entity Level}

\begin{table}[t]
    \centering
    \small
    \renewcommand{\arraystretch}{1.06}
    \caption{Entity-level proxy effectiveness for full-set visibility and NDCG ranking quality.}
    \label{tab:entity-proxy}
    \vspace{-2mm}

    \resizebox{\columnwidth}{!}{
    \begin{tabular}{llrrrrrrr}
        \toprule
        & & \multicolumn{3}{c}{Full-set visibility} & \multicolumn{4}{c}{NDCG} \\
        \cmidrule(lr){3-5}
        \cmidrule(lr){6-9}
        Setting & Proxy
        & Precision & Recall & F1
        & @1 & @3 & @5 & @10 \\
        \midrule
        5.3 & $G$
            & 25.9 & 19.1 & 19.4
            & 0.32 & 0.32 & 0.30 & 0.28 \\
        5.3 & $A_{\mathrm{cand}}$
            & 18.5 & 44.0 & 24.4
            & 0.34 & 0.31 & 0.30 & 0.31 \\
        5.3 & $A_{\mathrm{cite}}$
            & 28.7 & 30.2 & 27.5
            & 0.28 & 0.30 & 0.29 & 0.31 \\
        5.3 & $G\cup A_{\mathrm{cand}}$
            & 17.1 & 50.7 & 24.4
            & -- & -- & -- & -- \\
        \midrule
        5.5 & $G$
            & 22.0 & 18.6 & 18.2
            & 0.39 & 0.33 & 0.31 & 0.29 \\
        5.5 & $A_{\mathrm{cand}}$
            & 18.1 & 38.0 & 22.7
            & 0.36 & 0.34 & 0.32 & 0.33 \\
        5.5 & $A_{\mathrm{cite}}$
            & 30.1 & 26.4 & 25.5
            & 0.42 & 0.35 & 0.34 & 0.34 \\
        5.5 & $G\cup A_{\mathrm{cand}}$
            & 16.2 & 45.1 & 22.5
            & -- & -- & -- & -- \\
        \bottomrule
    \end{tabular}
    }
    \vspace{-2mm}
\end{table}

\stitle{{Responses API candidates achieve the highest entity recall}} As shown in Table~\ref{tab:entity-proxy}, the coverage--precision trade-off also holds at the entity level. Among individual proxies, $A_{\mathrm{cand}}$ achieves the highest Recall, at 44.0\% for \webthree\ and 38.0\% for \webfive, while $A_{\mathrm{cite}}$ achieves the highest Precision and F1. The union $G\cup A_{\mathrm{cand}}$ provides the broadest coverage, recovering 50.7\% and 45.1\% of cited entities, respectively. Its lower Precision, however, prevents a corresponding improvement in F1. Thus, Responses API candidates provide broader entity coverage, Responses API citations provide a more selective proxy, and Google Search adds further coverage when combined with the candidate proxy.

\begin{table*}[!t]
    \centering
    \small
    \setlength{\tabcolsep}{5pt}
    \renewcommand{\arraystretch}{1.10}
    \caption{Prediction features by signal family and granularity.}
    \label{tab:prediction-features}
    \vspace{-2mm}
    \begin{tabular}{llll}
        \toprule
        Category & Feature & Granularity & Definition \\
        \midrule

        Position/rank
        & \texttt{rrf\_score}
        & All levels
        & {Reciprocal-rank fusion of $G$ and $A_{\mathrm{cand}}$ ranks.}
        \\

        & \texttt{api\_best\_rank\_log}
        & URL/hostname
        & {Best $A_{\mathrm{cand}}$ rank across runs.}
        \\

        Exposure frequency
        & \texttt{api\_appear\_rate}
        & URL/hostname
        & {Cross-run $A_{\mathrm{cand}}$ appearance rate.}
        \\

        & \texttt{api\_exposure\_per\_run}
        & Entity
        & Entity exposure intensity $X_A(e)/20$.
        \\

        Conversion
        & \texttt{api\_citation\_given\_candidate}
        & URL/hostname
        & Citation-given-appearance rate.
        \\

        & \texttt{api\_citation\_rate}
        & Entity
        & Entity conversion ratio $C_A(e)/X_A(e)$.
        \\

        Support/consensus
        & \texttt{api\_url\_count\_log}
        & Hostname
        & {Distinct supporting $A_{\mathrm{cand}}$ URLs.}
        \\

        & \texttt{api\_url\_support\_log}
        & Entity
        & {Distinct supporting $A_{\mathrm{cand}}$ URLs $N_A(e)$.}
        \\

        & \texttt{source\_consensus}
        & Entity
        & {Entity agreement between $G$ and $A_{\mathrm{cand}}$.}
        \\

        Snippet/query
        & \texttt{snippet\_length\_log}
        & URL/hostname
        & Available snippet length.
        \\

        & Intent one-hot
        & All levels
        & Query intent.
        \\

        & \texttt{language\_chinese}
        & All levels
        & Query language.
        \\

        \bottomrule
    \end{tabular}
    \vspace{-2mm}
\end{table*}

\stitle{{Responses API citations achieve the highest entity NDCG for \webfive}} The ranking results differ between the two settings. For \webthree, no proxy consistently achieves the highest NDCG across the four cutoffs. For \webfive, $A_{\mathrm{cite}}$ achieves the highest NDCG at all four cutoffs, with NDCG@10 increasing from 0.311 for \webthree\ to 0.336 for \webfive. Its Precision also increases from 28.7\% to 30.1\%. In contrast, its Recall decreases from 30.2\% to 26.4\%. Recall also decreases for $G$, $A_{\mathrm{cand}}$, and $G\cup A_{\mathrm{cand}}$. Thus, Responses API citations better prioritize entities receiving stronger support from cited URLs in \webfive, while the external proxies recover a smaller share of cited entities overall. We evaluate a unified entity layer that includes brands under the extraction procedure in Section~\ref{sec:prelim}. These results therefore support entity-level visibility analysis but do not support separate conclusions for brands and other entity types. Even the union recovers only about half of the cited entities.

\section{Citation Prediction}
\label{sec:exp}

\subsection{Experiment Setup}

\stitle{Formulation}
We next test whether the external signals identified above can jointly distinguish cited from uncited candidates.
To do so, we formulate a binary classification task over the per-query candidate set $G\cup A_{\mathrm{cand}}$.
Each instance is a candidate-query pair at the URL, hostname, or entity level.
A URL is positive if it receives $y_W=1$ in at least one of the 20 ChatGPT Web runs.
A hostname is positive if any URL on that hostname is positive, and an entity is positive if $C_W(e)>0$.
All other items in $G\cup A_{\mathrm{cand}}$ are treated as negatives.
Cited items outside this candidate set are excluded rather than treated as negatives.
Thus, the task predicts citation conditional on external candidate discovery.
The predictor ranks only candidates in $G\cup A_{\mathrm{cand}}$, while the Recall of this candidate set reported in Section~\ref{sec:proxy} gives the maximum end-to-end Recall achievable under this candidate set.
ChatGPT Web is used only to construct labels and evaluate predictions.

\stitle{Features}
Table~\ref{tab:prediction-features} summarizes the prediction features.
All features are derived from $G$, $A_{\mathrm{cand}}$, $A_{\mathrm{cite}}$, or query metadata, and no ChatGPT Web retrieval signals are used as input.
We use rank, exposure frequency, conversion, multi-URL support, snippet, intent, and language signals, resulting in 9, 10, and 9 input dimensions for URL, hostname, and entity prediction, respectively.
Features prefixed with \texttt{api\_} use Responses API evidence from $A_{\mathrm{cand}}$ or $A_{\mathrm{cite}}$.
At the entity level, $X_A(e)$ and $C_A(e)$ are the exposure and citation aggregates defined in Section~\ref{sec:proxy}, while $N_A(e)$ counts the distinct $A_{\mathrm{cand}}$ URLs associated with entity $e$.
The suffix \texttt{\_log} denotes $\log(1+x)$.

\stitle{Models and baselines}
We compare a supervised XGBoost classifier~\cite{chen2016xgboost} with three rule-based rankings and a random baseline.
\textsc{Full-XGBoost} uses all features in Table~\ref{tab:prediction-features}.
\textsc{G-A-RRF} combines the $G$ and $A_{\mathrm{cand}}$ rankings defined in Section~\ref{sec:proxy} using reciprocal-rank fusion with $k_0=60$.
\textsc{Overall-RRF} additionally combines rankings induced by the remaining non-overlapping proxy features without reusing evidence already included in \textsc{G-A-RRF}.
\textsc{A-Citation-Only} ranks candidates using $A_{\mathrm{cite}}$ evidence, while \textsc{Random} assigns random scores within each query.
We train a separate XGBoost model for each ChatGPT Web setting and prediction granularity.
We use five-fold out-of-fold evaluation grouped by query and balanced by intent and language, so all candidates associated with the same query remain in the same fold.

\stitle{Metrics}
Our primary metric is query-macro ROC-AUC, and we additionally report Average Precision for the imbalanced labels.
We also report Precision, Recall, and F1 at $k\in\{1,3,5,10\}$.
Each metric is computed within each evaluable query and then macro-averaged.
We obtain 95\% confidence intervals from 2,000 query-level bootstrap samples.
Top-$k$ Recall is computed over cited items present in $G\cup A_{\mathrm{cand}}$.
The candidate-set Recall reported in Section~\ref{sec:proxy} separately measures cited items that never enter the prediction candidate set.
For pairwise method comparisons, we bootstrap query-level metric differences using the same 2,000 samples.

\subsection{Performance Evaluation}

\stitle{{\textsc{Full-XGBoost} leads five of six AUC comparisons}}
As shown in Table~\ref{tab:prediction-auc-ap}, \textsc{Full-XGBoost} achieves the highest AUC point estimate in five of the six setting-granularity combinations and ties with \textsc{Overall-RRF} for \webfive\ entities after rounding, with both reaching 0.742.
Its URL, hostname, and entity AUCs are 0.782, 0.763, and 0.743 for \webthree, and 0.779, 0.763, and 0.742 for \webfive, respectively.
Compared with \textsc{A-Citation-Only}, the AUC gains range from 0.068 to 0.110, and paired query-level bootstrap tests show positive AUC and AP differences in all six setting-granularity combinations.
\textsc{A-Citation-Only} uses only Responses API citation evidence, while \textsc{Full-XGBoost} additionally combines rank, exposure frequency, conversion, support, and snippet signals.
\textsc{Overall-RRF} reaches AUCs of 0.768, 0.726, and 0.742 for \webthree, and 0.772, 0.756, and 0.742 for \webfive\ at the URL, hostname, and entity levels, respectively.
For \webfive, it improves over \textsc{G-A-RRF} in both AUC and AP at all three granularities.
For \webthree, it improves both metrics for URLs and entities, while hostname AP increases despite a lower AUC of 0.726 compared with 0.737.
Thus, combining additional proxy signals improves the rule-based ranking in most settings, although fixed rank fusion does not consistently match the supervised predictor.

\begin{table}[t]
    \centering
    \small
    \setlength{\tabcolsep}{2.1pt}
    \renewcommand{\arraystretch}{1.06}
    \caption{Conditional citation prediction at the URL, hostname, and entity levels. Each cell reports query-macro ROC-AUC/AP; best values are bolded.}
    \vspace{-2mm}
    \label{tab:prediction-auc-ap}
    \begin{tabular}{@{\extracolsep{\fill}}llrrr@{}}
        \toprule
        Setting & Method & URL & Hostname & Entity \\
        \midrule
        5.3 & \textsc{Full-XGBoost}
            & \textbf{0.782}/0.286
            & \textbf{0.763}/\textbf{0.478}
            & \textbf{0.743}/\textbf{0.472} \\
        5.3 & \textsc{Overall-RRF}
            & 0.768/\textbf{0.294}
            & 0.726/0.467
            & 0.742/0.449 \\
        5.3 & \textsc{G-A-RRF}
            & 0.728/0.245
            & 0.737/0.430
            & 0.723/0.418 \\
        5.3 & \textsc{A-Citation-Only}
            & 0.675/0.237
            & 0.653/0.341
            & 0.671/0.336 \\
        5.3 & \textsc{Random}
            & 0.498/0.092
            & 0.501/0.181
            & 0.501/0.209 \\
        \midrule
        5.5 & \textsc{Full-XGBoost}
            & \textbf{0.779}/\textbf{0.334}
            & \textbf{0.763}/\textbf{0.503}
            & \textbf{0.742}/\textbf{0.474} \\
        5.5 & \textsc{Overall-RRF}
            & 0.772/0.313
            & 0.756/0.489
            & \textbf{0.742}/0.452 \\
        5.5 & \textsc{G-A-RRF}
            & 0.699/0.219
            & 0.716/0.409
            & 0.708/0.416 \\
        5.5 & \textsc{A-Citation-Only}
            & 0.699/0.283
            & 0.693/0.396
            & 0.674/0.336 \\
        5.5 & \textsc{Random}
            & 0.500/0.091
            & 0.501/0.175
            & 0.499/0.209 \\
        \bottomrule
    \end{tabular}
    \vspace{-2mm}
\end{table}

\stitle{{AUC remains stable across settings, while top-$k$ F1 improves at coarser levels}}
The \textsc{Full-XGBoost} AUC differs by at most 0.002 between \webthree\ and \webfive\ at each granularity.
AP increases from 0.286 to 0.334 for URLs and from 0.478 to 0.503 for hostnames, while remaining nearly unchanged for entities at 0.472 and 0.474.
AUC measures pairwise separability, whereas AP is more sensitive to positive prevalence and the ordering near the top.
The results therefore do not support a general claim that citations in \webfive\ are easier to predict.

\begin{figure}[t]
    \centering
    \begingroup
    \definecolor{prBlue}{HTML}{0072B2}
    \definecolor{prOrange}{HTML}{D55E00}
    \definecolor{prGreen}{HTML}{009E73}
    \definecolor{prPink}{HTML}{CC79A7}
    \definecolor{prGray}{HTML}{777777}

    \pgfplotsset{
        prediction pr axis/.style={
            width=2.2cm,
            height=1.6cm,
            scale only axis,
            xmin=0,
            xmax=0.5,
            ymin=0,
            ymax=0.9,
            xtick={0,0.25,0.5},
            xticklabels={0,.25,.5},
            ytick={0,0.3,0.6,0.9},
            tick label style={font=\footnotesize},
            tick style={line width=0.35pt},
            title style={font=\footnotesize, yshift=-1pt},
            axis line style={line width=0.45pt},
            axis lines=left,
            xmajorgrids=true,
            ymajorgrids=true,
            grid style={dashed, gray!22, line width=0.35pt},
            clip=true,
        },
        prediction pr xgb/.style={
            color=prBlue,
            solid,
            line width=0.75pt,
            mark=*,
            mark options={solid, fill=prBlue, draw=prBlue},
        },
        prediction pr overall/.style={
            color=prOrange,
            solid,
            line width=0.75pt,
            mark=square*,
            mark options={solid, fill=prOrange, draw=prOrange},
        },
        prediction pr ga/.style={
            color=prGreen,
            dashed,
            line width=0.75pt,
            mark=triangle*,
            mark options={solid, fill=white, draw=prGreen, line width=0.55pt},
        },
        prediction pr acite/.style={
            color=prPink,
            dash pattern=on 2.4pt off 0.9pt on 0.6pt off 0.9pt,
            line width=0.75pt,
            mark=diamond*,
            mark options={solid, fill=white, draw=prPink, line width=0.55pt},
        },
        prediction pr random/.style={
            color=prGray,
            dotted,
            line width=0.7pt,
            mark=x,
            mark options={solid, draw=prGray, line width=0.55pt},
        },
    }

    \newcommand{\predictionprplot}[5]{%
        \addplot[#1, no marks, forget plot] coordinates {#2 #3 #4 #5};%
        \addplot[#1, only marks, mark size=0.75pt, forget plot] coordinates {#2};%
        \addplot[#1, only marks, mark size=1.05pt, forget plot] coordinates {#3};%
        \addplot[#1, only marks, mark size=1.35pt, forget plot] coordinates {#4};%
        \addplot[#1, only marks, mark size=1.70pt, forget plot] coordinates {#5};%
    }
    \newcommand{\predictionprcontours}{%
        \addplot[gray!38, dashed, line width=0.35pt, domain=0.053:0.5, samples=36, forget plot] {0.1*x/(2*x-0.1)};%
        \addplot[gray!38, dashed, line width=0.35pt, domain=0.113:0.5, samples=36, forget plot] {0.2*x/(2*x-0.2)};%
        \addplot[gray!38, dashed, line width=0.35pt, domain=0.180:0.5, samples=36, forget plot] {0.3*x/(2*x-0.3)};%
        \addplot[gray!38, dashed, line width=0.35pt, domain=0.258:0.5, samples=36, forget plot] {0.4*x/(2*x-0.4)};%
    }
    \newcommand{\predictionprfivlabels}{%
        \node[anchor=east, font=\tiny, text=gray!72] at (axis cs:0.485,0.410) {$F_1=0.4$};%
        \node[anchor=east, font=\tiny, text=gray!72] at (axis cs:0.485,0.244) {$F_1=0.3$};%
        \node[anchor=east, font=\tiny, text=gray!72] at (axis cs:0.485,0.134) {$F_1=0.2$};%
        \node[anchor=east, font=\tiny, text=gray!72] at (axis cs:0.485,0.057) {$F_1=0.1$};%
    }
    \begin{tikzpicture}
        \begin{axis}[
            hide axis,
            width=1cm,
            height=0.6cm,
            scale only axis,
            xmin=0,
            xmax=1,
            ymin=0,
            ymax=1,
            legend columns=3,
            legend style={
                draw=none,
                font=\footnotesize,
                column sep=0.55em,
                /tikz/every even column/.append style={column sep=0.25em},
            },
        ]
            \addlegendimage{prediction pr xgb, mark size=1.25pt}
            \addlegendentry{Full XGBoost}
            \addlegendimage{prediction pr overall, mark size=1.25pt}
            \addlegendentry{Overall RRF}
            \addlegendimage{prediction pr ga, mark size=1.25pt}
            \addlegendentry{G-A RRF}
            \addlegendimage{prediction pr acite, mark size=1.25pt}
            \addlegendentry{A-citation only}
            \addlegendimage{prediction pr random, mark size=1.25pt}
            \addlegendentry{Random}
        \end{axis}
    \end{tikzpicture}\\[-0.2em]

    {\small\bfseries GPT-5.3 Instant}\\[-0.35em]

    \begin{tikzpicture}
        \begin{axis}[
            prediction pr axis,
            name=prA,
            title={URL},
            ylabel={Precision},
            ylabel style={font=\scriptsize},
        ]
            \predictionprcontours
            \predictionprplot{prediction pr xgb}{(0.0500,0.3506)}{(0.1407,0.3247)}{(0.1945,0.2779)}{(0.3349,0.2429)}
            \predictionprplot{prediction pr overall}{(0.0580,0.4026)}{(0.1640,0.3636)}{(0.2445,0.3403)}{(0.3232,0.2377)}
            \predictionprplot{prediction pr ga}{(0.0494,0.3506)}{(0.1266,0.3095)}{(0.1658,0.2571)}{(0.2683,0.2117)}
            \predictionprplot{prediction pr acite}{(0.0705,0.3506)}{(0.1427,0.3038)}{(0.1995,0.2755)}{(0.2866,0.2120)}
            \predictionprplot{prediction pr random}{(0.0080,0.0605)}{(0.0220,0.0588)}{(0.0380,0.0598)}{(0.0741,0.0595)}
        \end{axis}

        \begin{axis}[prediction pr axis, name=prB, at={(prA.south east)}, anchor=south west, xshift=0.22cm, title={Hostname}, yticklabel=\empty]
            \predictionprcontours
            \predictionprplot{prediction pr xgb}{(0.0792,0.6875)}{(0.2040,0.6083)}{(0.3073,0.5425)}{(0.4523,0.4163)}
            \predictionprplot{prediction pr overall}{(0.0797,0.6750)}{(0.2128,0.6292)}{(0.3015,0.5425)}{(0.4525,0.4088)}
            \predictionprplot{prediction pr ga}{(0.0759,0.6562)}{(0.1802,0.5417)}{(0.2511,0.4688)}{(0.3924,0.3669)}
            \predictionprplot{prediction pr acite}{(0.0791,0.6188)}{(0.1730,0.4836)}{(0.2407,0.4184)}{(0.3486,0.3149)}
            \predictionprplot{prediction pr random}{(0.0148,0.1355)}{(0.0439,0.1352)}{(0.0723,0.1336)}{(0.1452,0.1339)}
        \end{axis}

        \begin{axis}[prediction pr axis, name=prC, at={(prB.south east)}, anchor=south west, xshift=0.22cm, title={Entity}, yticklabel=\empty]
            \predictionprcontours
            \predictionprplot{prediction pr xgb}{(0.0544,0.6750)}{(0.1436,0.6208)}{(0.2097,0.5800)}{(0.3503,0.4956)}
            \predictionprplot{prediction pr overall}{(0.0511,0.6875)}{(0.1351,0.5792)}{(0.1992,0.5325)}{(0.3175,0.4494)}
            \predictionprplot{prediction pr ga}{(0.0451,0.6250)}{(0.1259,0.5729)}{(0.1880,0.5238)}{(0.3039,0.4281)}
            \predictionprplot{prediction pr acite}{(0.0316,0.3529)}{(0.0959,0.3744)}{(0.1424,0.3621)}{(0.2792,0.3651)}
            \predictionprplot{prediction pr random}{(0.0117,0.1724)}{(0.0354,0.1737)}{(0.0594,0.1739)}{(0.1182,0.1733)}
            \predictionprfivlabels
        \end{axis}

    \end{tikzpicture}\\[-0.1em]

    {\small\bfseries GPT-5.5 Instant}\\[-0.35em]

    \begin{tikzpicture}
        \begin{axis}[
            prediction pr axis,
            name=prD,
            title={URL},
            ylabel={Precision},
            ylabel style={font=\scriptsize},
        ]
            \predictionprcontours
            \predictionprplot{prediction pr xgb}{(0.0683,0.4744)}{(0.1723,0.4487)}{(0.2452,0.3897)}{(0.3668,0.2910)}
            \predictionprplot{prediction pr overall}{(0.0630,0.5256)}{(0.1589,0.4060)}{(0.2295,0.3692)}{(0.3583,0.2821)}
            \predictionprplot{prediction pr ga}{(0.0464,0.3782)}{(0.1106,0.2714)}{(0.1699,0.2526)}{(0.2527,0.1904)}
            \predictionprplot{prediction pr acite}{(0.0825,0.5545)}{(0.1721,0.4269)}{(0.2380,0.3691)}{(0.3327,0.2756)}
            \predictionprplot{prediction pr random}{(0.0083,0.0617)}{(0.0239,0.0605)}{(0.0401,0.0607)}{(0.0786,0.0599)}
        \end{axis}

        \begin{axis}[prediction pr axis, name=prE, at={(prD.south east)}, anchor=south west, xshift=0.22cm, title={Hostname}, yticklabel=\empty]
            \predictionprcontours
            \predictionprplot{prediction pr xgb}{(0.1048,0.8250)}{(0.2248,0.6625)}{(0.3279,0.5875)}{(0.4657,0.4300)}
            \predictionprplot{prediction pr overall}{(0.0983,0.7875)}{(0.2046,0.6083)}{(0.3060,0.5600)}{(0.4710,0.4450)}
            \predictionprplot{prediction pr ga}{(0.0856,0.6562)}{(0.1868,0.5500)}{(0.2584,0.4738)}{(0.3820,0.3619)}
            \predictionprplot{prediction pr acite}{(0.0987,0.7375)}{(0.2205,0.6100)}{(0.2955,0.5147)}{(0.4088,0.3731)}
            \predictionprplot{prediction pr random}{(0.0137,0.1247)}{(0.0424,0.1292)}{(0.0713,0.1301)}{(0.1411,0.1288)}
        \end{axis}

        \begin{axis}[prediction pr axis, name=prF, at={(prE.south east)}, anchor=south west, xshift=0.22cm, title={Entity}, yticklabel=\empty]
            \predictionprcontours
            \predictionprplot{prediction pr xgb}{(0.0541,0.7115)}{(0.1456,0.6410)}{(0.2246,0.5872)}{(0.3814,0.5038)}
            \predictionprplot{prediction pr overall}{(0.0497,0.6410)}{(0.1379,0.5983)}{(0.2199,0.5538)}{(0.3701,0.4679)}
            \predictionprplot{prediction pr ga}{(0.0496,0.6218)}{(0.1350,0.5620)}{(0.2092,0.5269)}{(0.3365,0.4250)}
            \predictionprplot{prediction pr acite}{(0.0354,0.4610)}{(0.1023,0.4339)}{(0.1641,0.4114)}{(0.2934,0.3656)}
            \predictionprplot{prediction pr random}{(0.0135,0.1640)}{(0.0413,0.1639)}{(0.0692,0.1652)}{(0.1405,0.1670)}
            \predictionprfivlabels
        \end{axis}
    \end{tikzpicture}\\[-0.45em]
    \makebox[\columnwidth][c]{\scriptsize Recall}%
    \makebox[0pt][r]{\raisebox{0pt}[0pt][0pt]{\footnotesize Marker size: $k=1,3,5,10$}\hspace{0.6em}}
    \endgroup
    \caption{{Top-$k$ precision and recall across ChatGPT Web settings at the URL, hostname, and entity levels. Gray dashed curves show equal-F1 contours.}}
    \label{fig:prediction-topk}
    \vspace{-2mm}
\end{figure}
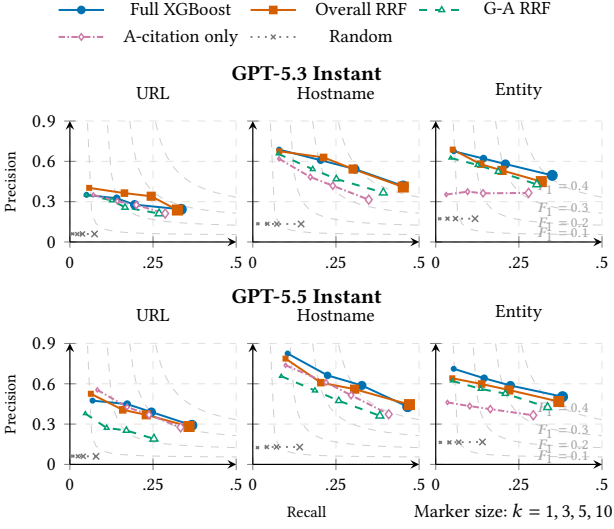

{As shown in Figure~\ref{fig:prediction-topk}, prediction at the hostname and entity levels produces stronger top-$k$ shortlists than exact-URL prediction.}
At $k=10$, \textsc{Full-XGBoost} reaches F1 values of 0.257, 0.410, and 0.369 for \webthree, and 0.294, 0.420, and 0.397 for \webfive\ at the URL, hostname, and entity levels, respectively.
\textsc{Overall-RRF} exceeds \textsc{Full-XGBoost} for \webfive\ hostnames at $k=10$, with F1 values of 0.433 and 0.420, respectively, but remains behind \textsc{Full-XGBoost} for entity prediction under both settings.
Hostname and entity aggregation combines evidence across multiple URLs and reduces the sparsity of exact-URL labels, which helps explain their stronger top-$k$ performance.

\stitle{{External candidate coverage caps exact-URL recall}}
The prediction results above are conditional on the external candidate set $G\cup A_{\mathrm{cand}}$.
For \webthree, this set covers 28.3\%, 50.3\%, and 50.7\% of cited URLs, hostnames, and entities, respectively.
The corresponding coverage for \webfive\ is 26.2\%, 48.7\%, and 45.1\%.
{Thus, even perfect ranking within the candidate set can recover only 26.2\%--28.3\% of all cited URLs.}
Cited URLs absent from both $G$ and $A_{\mathrm{cand}}$ never enter the prediction task.
End-to-end citation prediction therefore depends on broader candidate discovery in addition to more accurate conditional ranking.

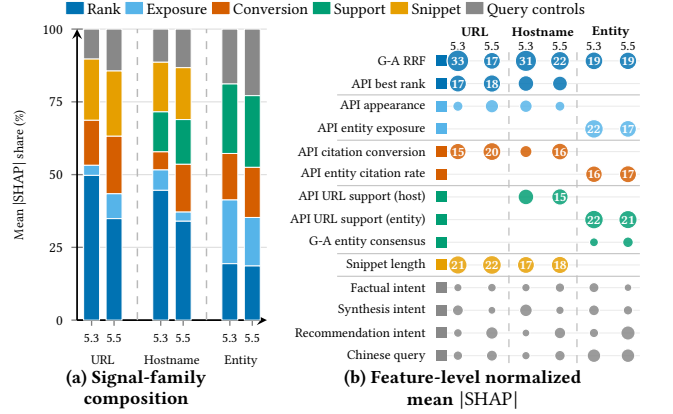
\begin{figure}[t]
    \centering
    \begingroup
    \definecolor{shapRank}{HTML}{0072B2}
    \definecolor{shapFrequency}{HTML}{56B4E9}
    \definecolor{shapConversion}{HTML}{D55E00}
    \definecolor{shapSupport}{HTML}{009E73}
    \definecolor{shapSnippet}{HTML}{E69F00}
    \definecolor{shapControls}{HTML}{888888}

    \newcommand{\shaplegenditem}[2]{%
        \tikz[baseline=-0.55ex]{\node[fill=#1, draw=white, line width=0.2pt, minimum width=7pt, minimum height=5pt, inner sep=0pt] {};}%
        \hspace{0.12em}{\scriptsize #2}%
    }
    \newcommand{\shapbubble}[5]{%
        \pgfmathsetmacro{\shapradius}{0.9 + 3.2 * sqrt((#3) / 33)}%
        \filldraw[fill=#4, fill opacity=0.84, draw=white, line width=0.45pt]
            (#1,#2) circle[radius=\shapradius pt];%
        \node[font=\tiny\bfseries, text=white]
            at (#1,#2) {#5};%
    }
    \newcommand{\shaprowlabel}[4]{%
        \node[anchor=east, font=\tiny]
            at (-0.70,#1) {#2};%
        \node[fill=#3, minimum width=3.8pt, minimum height=3.8pt, inner sep=0pt]
            at (-0.49,#1) {};%
        \draw[gray!16, line width=0.4pt] (-0.32,#1) -- (5.35,#1);%
        #4%
    }

    \makebox[\columnwidth][c]{%
        \shaplegenditem{shapRank}{Rank}\hspace{0.16em}%
        \shaplegenditem{shapFrequency}{Exposure}\hspace{0.16em}%
        \shaplegenditem{shapConversion}{Conversion}\hspace{0.16em}%
        \shaplegenditem{shapSupport}{Support}\hspace{0.16em}%
        \shaplegenditem{shapSnippet}{Snippet}\hspace{0.16em}%
        \shaplegenditem{shapControls}{Query controls}%
    }\par\vspace{-0.15em}

    \begin{minipage}[b]{0.41\columnwidth}
    \centering
    \begin{tikzpicture}
        \begin{axis}[
            width=2.51cm,
            height=3.85cm,
            scale only axis,
            ybar stacked,
            bar width=6pt,
            ymin=0,
            ymax=100,
            xmin=-0.42,
            xmax=5.07,
            ytick={0,25,50,75,100},
            ylabel={Mean $|\mathrm{SHAP}|$ share (\%)},
            ylabel style={font=\tiny},
            xtick={0,0.65,2,2.65,4,4.65},
            xticklabels={5.3,5.5,5.3,5.5,5.3,5.5},
            xticklabel style={font=\tiny},
            extra x ticks={0.325,2.325,4.325},
            extra x tick labels={URL,Hostname,Entity},
            extra x tick style={
                tick style={draw=none},
                tick label style={font=\tiny\bfseries, yshift=-2.0ex},
            },
            tick label style={font=\tiny},
            tick style={line width=0.5pt},
            axis lines=left,
            axis line style={line width=0.65pt},
            ymajorgrids=true,
            grid style={gray!24, line width=0.45pt},
            clip=false,
        ]
            \addplot[fill=shapRank, draw=white, line width=0.35pt] coordinates {
                (0,49.728) (0.65,34.867) (2,44.596) (2.65,33.973) (4,19.384) (4.65,18.600)
            };
            \addplot[fill=shapFrequency, draw=white, line width=0.35pt] coordinates {
                (0,3.464) (0.65,8.545) (2,7.003) (2.65,3.150) (4,21.898) (4.65,16.669)
            };
            \addplot[fill=shapConversion, draw=white, line width=0.35pt] coordinates {
                (0,15.466) (0.65,19.767) (2,6.213) (2.65,16.413) (4,15.978) (4.65,17.220)
            };
            \addplot[fill=shapSupport, draw=white, line width=0.35pt] coordinates {
                (0,0.000) (0.65,0.000) (2,13.740) (2.65,15.367) (4,23.874) (4.65,24.632)
            };
            \addplot[fill=shapSnippet, draw=white, line width=0.35pt] coordinates {
                (0,21.039) (0.65,22.420) (2,17.055) (2.65,17.833) (4,0.000) (4.65,0.000)
            };
            \addplot[fill=shapControls, draw=white, line width=0.35pt] coordinates {
                (0,10.303) (0.65,14.401) (2,11.393) (2.65,13.264) (4,18.866) (4.65,22.880)
            };
            \draw[gray!55, dashed, line width=0.55pt] (axis cs:1.325,0) -- (axis cs:1.325,100);
            \draw[gray!55, dashed, line width=0.55pt] (axis cs:3.325,0) -- (axis cs:3.325,100);
        \end{axis}
    \end{tikzpicture}

    \par\vspace{-0.35em}
    \makebox[\linewidth][c]{\parbox{\linewidth}{\centering\footnotesize\bfseries (a) Signal-family\\composition}}
    \end{minipage}\hfill
    \begin{minipage}[b]{0.57\columnwidth}
    \centering
    \begin{tikzpicture}[x=0.45cm,y=0.30cm]
        \node[font=\tiny\bfseries] at (0.5,1.28) {URL};
        \node[font=\tiny\bfseries] at (2.5,1.28) {Hostname};
        \node[font=\tiny\bfseries] at (4.5,1.28) {Entity};
        \foreach \x/\model in {0/5.3,1/5.5,2/5.3,3/5.5,4/5.3,5/5.5} {
            \node[font=\tiny] at (\x,0.67) {\model};
        }

        \shaprowlabel{0}{G-A RRF}{shapRank}{
            \shapbubble{0}{0}{32.527}{shapRank}{33}
            \shapbubble{1}{0}{16.766}{shapRank}{17}
            \shapbubble{2}{0}{30.931}{shapRank}{31}
            \shapbubble{3}{0}{22.380}{shapRank}{22}
            \shapbubble{4}{0}{19.384}{shapRank}{19}
            \shapbubble{5}{0}{18.600}{shapRank}{19}
        }
        \shaprowlabel{-1}{API best rank}{shapRank}{
            \shapbubble{0}{-1}{17.201}{shapRank}{17}
            \shapbubble{1}{-1}{18.101}{shapRank}{18}
            \shapbubble{2}{-1}{13.665}{shapRank}{}
            \shapbubble{3}{-1}{11.594}{shapRank}{}
        }
        \shaprowlabel{-2}{API appearance}{shapFrequency}{
            \shapbubble{0}{-2}{3.464}{shapFrequency}{}
            \shapbubble{1}{-2}{8.545}{shapFrequency}{}
            \shapbubble{2}{-2}{7.003}{shapFrequency}{}
            \shapbubble{3}{-2}{3.150}{shapFrequency}{}
        }
        \shaprowlabel{-3}{API entity exposure}{shapFrequency}{
            \shapbubble{4}{-3}{21.898}{shapFrequency}{22}
            \shapbubble{5}{-3}{16.669}{shapFrequency}{17}
        }
        \shaprowlabel{-4}{API citation conversion}{shapConversion}{
            \shapbubble{0}{-4}{15.466}{shapConversion}{15}
            \shapbubble{1}{-4}{19.767}{shapConversion}{20}
            \shapbubble{2}{-4}{6.213}{shapConversion}{}
            \shapbubble{3}{-4}{16.413}{shapConversion}{16}
        }
        \shaprowlabel{-5}{API entity citation rate}{shapConversion}{
            \shapbubble{4}{-5}{15.978}{shapConversion}{16}
            \shapbubble{5}{-5}{17.220}{shapConversion}{17}
        }
        \shaprowlabel{-6}{API URL support (host)}{shapSupport}{
            \shapbubble{2}{-6}{13.740}{shapSupport}{}
            \shapbubble{3}{-6}{15.367}{shapSupport}{15}
        }
        \shaprowlabel{-7}{API URL support (entity)}{shapSupport}{
            \shapbubble{4}{-7}{21.896}{shapSupport}{22}
            \shapbubble{5}{-7}{20.697}{shapSupport}{21}
        }
        \shaprowlabel{-8}{G-A entity consensus}{shapSupport}{
            \shapbubble{4}{-8}{1.977}{shapSupport}{}
            \shapbubble{5}{-8}{3.935}{shapSupport}{}
        }
        \shaprowlabel{-9}{Snippet length}{shapSnippet}{
            \shapbubble{0}{-9}{21.039}{shapSnippet}{21}
            \shapbubble{1}{-9}{22.420}{shapSnippet}{22}
            \shapbubble{2}{-9}{17.055}{shapSnippet}{17}
            \shapbubble{3}{-9}{17.833}{shapSnippet}{18}
        }
        \shaprowlabel{-10}{Factual intent}{shapControls}{
            \shapbubble{0}{-10}{1.431}{shapControls}{}
            \shapbubble{1}{-10}{1.789}{shapControls}{}
            \shapbubble{2}{-10}{1.561}{shapControls}{}
            \shapbubble{3}{-10}{2.555}{shapControls}{}
            \shapbubble{4}{-10}{3.272}{shapControls}{}
            \shapbubble{5}{-10}{1.028}{shapControls}{}
        }
        \shaprowlabel{-11}{Synthesis intent}{shapControls}{
            \shapbubble{0}{-11}{4.412}{shapControls}{}
            \shapbubble{1}{-11}{1.188}{shapControls}{}
            \shapbubble{2}{-11}{7.166}{shapControls}{}
            \shapbubble{3}{-11}{1.576}{shapControls}{}
            \shapbubble{4}{-11}{4.201}{shapControls}{}
            \shapbubble{5}{-11}{2.529}{shapControls}{}
        }
        \shaprowlabel{-12}{Recommendation intent}{shapControls}{
            \shapbubble{0}{-12}{1.833}{shapControls}{}
            \shapbubble{1}{-12}{6.880}{shapControls}{}
            \shapbubble{2}{-12}{1.122}{shapControls}{}
            \shapbubble{3}{-12}{5.290}{shapControls}{}
            \shapbubble{4}{-12}{2.430}{shapControls}{}
            \shapbubble{5}{-12}{10.259}{shapControls}{}
        }
        \shaprowlabel{-13}{Chinese query}{shapControls}{
            \shapbubble{0}{-13}{2.627}{shapControls}{}
            \shapbubble{1}{-13}{4.544}{shapControls}{}
            \shapbubble{2}{-13}{1.544}{shapControls}{}
            \shapbubble{3}{-13}{3.843}{shapControls}{}
            \shapbubble{4}{-13}{8.963}{shapControls}{}
            \shapbubble{5}{-13}{9.064}{shapControls}{}
        }

        \foreach \y in {-1.5,-3.5,-5.5,-8.5,-9.5} {
            \draw[gray!44, line width=0.45pt] (-3.55,\y) -- (5.35,\y);
        }
        \draw[gray!48, dashed, line width=0.5pt] (1.5,0.38) -- (1.5,-13.38);
        \draw[gray!48, dashed, line width=0.5pt] (3.5,0.38) -- (3.5,-13.38);
    \end{tikzpicture}
    \par\vspace{-0.15em}
    \makebox[\linewidth][c]{\parbox{\linewidth}{\centering\footnotesize\bfseries (b) Feature-level normalized\\mean $|\mathrm{SHAP}|$}}
    \end{minipage}
    \endgroup
    \caption{Feature importance of \textsc{Full-XGBoost}. Panel (a) shows signal-family shares. In panel (b), bubble area shows feature shares, labels mark shares of at least 15\%, and blank cells indicate unused features.}
    \vspace{-2mm}
    \label{fig:prediction-shap}
\end{figure}

\subsection{Importance Analysis}

We use TreeSHAP to quantify feature contributions to the fitted XGBoost predictors.
We compute TreeSHAP values on up to 2,000 held-out candidate rows from each cross-validation fold and aggregate their absolute magnitudes across folds.
Figure~\ref{fig:prediction-shap} reports the resulting importance at the signal-family and individual-feature levels.
For Figure~\ref{fig:prediction-shap}(a), we sum the mean absolute TreeSHAP values within each signal family and normalize the totals within each setting and granularity.
Panel (b) applies the same normalization to individual features.

\stitle{Source-side signals account for most of the predictive importance}
As shown in Figure~\ref{fig:prediction-shap}(a), rank, exposure frequency, conversion, support, and snippet signals derived from $G$, $A_{\mathrm{cand}}$, and $A_{\mathrm{cite}}$ account for 77.1\%--89.7\% of normalized importance, while intent and language account for the remainder.
Rank is the largest signal family for URL and hostname prediction, contributing 49.7\% and 44.6\% for \webthree, and 34.9\% and 34.0\% for \webfive, respectively.
This is consistent with the competitive performance of the rank-fusion baselines in Table~\ref{tab:prediction-auc-ap}, while the supervised predictor additionally uses exposure frequency, conversion, support, and snippet signals.

\stitle{{Rank importance decreases in \webfive}}
At the URL level, the rank share decreases from 49.7\% for \webthree\ to 34.9\% for \webfive, while exposure frequency and conversion jointly increase from 19.0\% to 28.3\%.
The same pattern appears at the hostname level, where rank decreases from 44.6\% to 34.0\%, while conversion and multi-URL support increase from 19.9\% to 31.8\%.
Snippet importance remains similar between the two settings.
This pattern is consistent with the stronger performance of $A_{\mathrm{cite}}$ as a proxy for \webfive\ reported in Section~\ref{sec:proxy}.
\stitle{Multi-URL support becomes more important at the hostname and entity levels}
Figure~\ref{fig:prediction-shap}(b) shows that exact-URL prediction relies mainly on rank, snippet, and citation-conversion signals.
At the hostname and entity levels, evidence from multiple URLs becomes more important.
Distinct API URL support contributes 13.7\% and 15.4\% of hostname importance for \webthree\ and \webfive, respectively, while support is the largest entity-level family at 23.9\% and 24.6\%.
Thus, aggregation increases the importance of repeated support across multiple URLs.

These importance values describe associations within the fitted predictors rather than causal effects, so changing an individual feature does not necessarily increase the probability of being cited.

\section{Generative Engine Optimization Implications}
\label{sec:geo}

Generative engine optimization (GEO)~\cite{aggarwal2024geo} aims to improve content visibility in generative-engine responses, where visibility depends not only on retrieval and ranking but also on citation selection. Our study does not directly evaluate GEO strategies. Instead, our measurements provide several implications for monitoring and improving visibility in ChatGPT Web. Recall from Section~\ref{sec:prelim} that our citation definition may include supporting webpages not separately displayed to users. Hence, these implications concern the broader reference-selection process rather than only user-visible citations.

\stitle{Retrieval visibility does not guarantee citation}
Table~\ref{tab:channel-proxy-overall} shows that Google Search covers only part of the sources cited by ChatGPT Web.
More importantly, the difference between $A_{\mathrm{cand}}$ and $A_{\mathrm{cite}}$ separates candidate exposure from citation selection.
A source may therefore appear in external search or retrieval results without being cited by ChatGPT Web, let alone appearing as a displayed citation.
GEO should distinguish retrieval visibility from citation visibility rather than treat search ranking as the final objective.

\stitle{GEO monitoring should account for rewritten queries and repeated runs}
ChatGPT frequently reformulates the original user query before retrieval, with the observed rewriting behavior differing across model settings (Table~\ref{tab:query-rewriting-overview}).
GEO monitoring should therefore consider the executed queries rather than only the user-facing query.
Repeated runs are also necessary because candidate sets vary substantially across responses, while candidates retrieved more often across runs are more likely to be cited (Figures~\ref{fig:candidate-stability} and~\ref{fig:recurrence-citation}).
A single response may therefore provide an incomplete view of source visibility.
However, repeated measurements may also trigger silent model fallback.
In our collection, two sessions switched to fallback models after 14 and 16 successful target-model runs.
GEO monitoring should therefore verify the serving model for each response to avoid unknowingly measuring a fallback model.

\stitle{External proxies are more informative at the hostname and entity levels than at the exact-URL level}
The OpenAI Responses API provides an external signal for GEO monitoring but does not reproduce ChatGPT Web exactly.
Among individual proxies, Responses API candidates achieve higher recall, while Responses API citations achieve higher precision and ranking quality.
Proxy effectiveness is also consistently higher at the hostname and entity levels than at the exact-URL level (Tables~\ref{tab:channel-proxy-overall} and~\ref{tab:entity-proxy}).
The same pattern appears across repeated ChatGPT Web runs, where hostnames and entities are more stable than exact URLs.
GEO monitoring may therefore benefit from tracking hostname- and entity-level presence in addition to individual webpages.
These results are observational and do not imply that publishing more pages or making a source appear more often in retrieval will increase its likelihood of being cited.

\stitle{GEO strategies should be re-evaluated when the model setting or serving environment changes}
Query rewriting, proxy effectiveness, and predictive signals differ across the observed ChatGPT Web settings.
Figure~\ref{fig:prediction-shap}, for example, shows that the relative importance of rank, exposure frequency, conversion, and support signals differs between \webthree\ and \webfive.
These observations do not isolate the effect of a model update, but they show that signals measured under one setting may not transfer unchanged to another.
GEO monitoring and optimization strategies should therefore be re-evaluated when the deployed model setting or serving environment changes.

\section{Additional Related Works}
\label{sec:related}

\stitle{Generative engine optimization}
The GEO framework~\citep{aggarwal2024geo,c-seo} formalizes content optimization for generative engines that retrieve, synthesize, and selectively cite webpages.
Subsequent work improves source visibility through document rewriting, structural modification, or adversarial intervention~\citep{ragdoll,black-hat-geo,tang2025stealthrank,DBLP:conf/emnlp/QianZJJW25,jin2026controlling,chen2026llmseo}.
For example, \citet{yu2026structural} study how document structure affects citation outcomes, while \citet{tian2026diagnosing} diagnose and repair citation failures through targeted rewriting.
Other methods study post-retrieval interventions on context use or citation generation~\citep{tan2025pear,yu2026c2cite}.
These studies primarily ask how to improve visibility through controlled interventions.
In contrast, we measure reference selection in ChatGPT Web, including the supporting webpages attached to selected references, without modifying documents or the model.
\citet{martinez2026optimizing} similarly models generative visibility as a multi-stage process, but does not observe the internal retrieval-to-citation process of a production engine.
We instead use SSE traces to observe executed queries, retrieved candidates, selected references, and supporting webpages across multiple ChatGPT Web model settings.

\stitle{Measurement studies of AI search systems}
Prior measurement studies examine source selection, citation behavior, and stability across generative search systems~\citep{zhang2026measurement,kirsten2025characterizing,yang2025news,huang2026answerbubbles}.
They report differences in citation selection, source diversity, concentration, and cross-system information exposure.
Closely related, \citet{grossman2026disrupts} compares Google Search, Google AI Overviews, and Gemini 2.5 Flash, finding low URL overlap across products and source instability under repeated runs and minor query changes.
These studies mainly compare final outputs across systems or products.
We instead observe the candidate-to-citation transition within ChatGPT Web and compare it across multiple model settings.

\section{Conclusion}
This work presents a multi-setting observational study of how ChatGPT Web rewrites queries, retrieves candidate webpages, and selects references across four settings from ChatGPT 5.2 to 5.5.
We find that retrieval behavior differs across settings and that candidates retrieved more often across repeated runs are more likely to be cited.
Google Search and the OpenAI Responses API only partially reproduce ChatGPT Web, with stronger results at the hostname and entity levels than at the exact-URL level, while their signals can predict which sources ChatGPT Web cites.
These findings suggest that GEO should distinguish retrieval visibility from citation visibility, account for rewritten queries and repeated runs, and consider hostname- and entity-level presence in addition to individual webpages.
Future work will extend this study to more generative engines and model settings.


\section{Ethical Considerations}
All queries used in this study come from the publicly available Search Arena dataset~\cite{miroyan2025search}, and our analysis considers only publicly accessible webpages and ChatGPT Web outputs.
The SSE traces are collected only from network responses returned to our own authenticated ChatGPT Web sessions.
We do not access other users' conversations, use private user data, or attempt to identify individuals.
We also pace repeated requests to limit their impact on the service.
Before releasing the collected traces, we remove session identifiers, request identifiers, device metadata, and other information unrelated to the measurements reported in this paper, and pseudonymize identifiers that are necessary for the analysis.

\bibliographystyle{ACM-Reference-Format}
\bibliography{refs}


\end{document}